\documentclass[a4paper,11pt]{article}
\pdfoutput=1 

\usepackage{jheppub} 
\usepackage[T1]{fontenc} 
\usepackage{esvect,bm,mathrsfs,makecell,graphicx,color,subcaption}
\usepackage{booktabs}
\usepackage[utf8]{inputenc}
\usepackage{slashed}
\usepackage{color}

\newcommand\ci{\perp\!\!\!\perp}

\title{Relativistic calculations of $R(D^{(*)})$, $R(D^{(*)}_s)$, $R(\eta_c)$ and $R(J/\psi)$}

\author[1]{Tian Zhou\note{tianzhou@hit.edu.cn.},}
\author[2]{Tianhong Wang\note{Corresponding author. thwang@hit.edu.cn.}}
\author[3]{, Yue Jiang\note{jiangure@hit.edu.cn},}
\author[4]{Xiao-Ze Tan\note{xz.tan@hit.edu.cn},}
\author[5]{Geng Li\note{karlisle@hit.edu.cn},}
\author[6]{Guo-Li Wang\note{glwang@hit.edu.cn}}

\affiliation[]{School of Physics, Harbin Institute of Technology, Harbin 150001, China}

\abstract{
Recently, the deviation of the ratios $R(D)$, $R(D^{*})$ and $R(J/\psi)$ have been found between experimental data and the Standard Model predictions, which may be the hint of New Physics. In this work, we calculate these ratios within the Standard Model by using the improved instantaneous Bethe-Salpeter method. The emphasis is pad to the relativistic correction of the form factors. The results are $R(D)=0.312 ^{+0.006}_{-0.007}$, $R(D^*)= 0.249^{+0.001}_{-0.002}$, $R(D_s)=0.320 ^{+0.009}_{-0.009}$, $R(D^*_s)=0.251 ^{+0.002}_{-0.003}$, $R(\eta_c)=0.384 ^{+0.032}_{-0.042}$, and $R(J/\psi)=0.267 ^{+0.009}_{-0.011}$, which are consistent with predictions of other models and the experimental data. The semileptonic decay rates and corresponding form factors at zero recoil are also given.}

\begin{document}
\maketitle
\flushbottom

\section{Introduction}
\label{sec:intro}

As we all believe, the Standard Model (SM) is not a perfect theory especially at higher scale, so it is significant to test SM precisely to search the new physics (NP) beyond SM\cite{Li:2018lxi}. Recently, several experiments reported a few anomalous results of $R(D^{(*)})$ and $R(J/\psi)$, which are defined as
\begin{equation}
R\left(D^{(*)}\right)=\frac{B r\left(B \rightarrow D^{(*)} \tau \nu\right)}{\operatorname{Br}\left(B \rightarrow D^{(*)} \ell \nu\right)},
\end{equation}
and
\begin{equation}
R(J / \psi)=\frac{B r\left(B_{c} \rightarrow J / \psi \tau \nu\right)}{\operatorname{Br}\left(B_{c} \rightarrow J / \psi \mu \nu\right)},
\end{equation}
respectively. The research of $R(D^{(*)})$ and $R(J/\psi)$ has become interesting, because people believe these quantities can be used to explore NP~\cite{Li:2018lxi,Graverini:2018riw,Amhis:2016xyh,Fajfer:2012jt,Sahoo:2016pet,Fajfer:2012vx,Nierste:2008qe,Datta:2012qk,Becirevic:2012jf,Crivellin:2012ye,Bauer:2015knc,Dorsner:2016wpm,Celis:2016azn}. Moreover, as the Cabibbo-Kobayashi-Maskawa (CKM) matrix element $V_{cb}$ contained in the branching fractions canceled each other out, the uncertainties originate form them are reduced.

These ratios have been measured by BaBar \cite{Lees:2012xj,Lees:2013uzd}, Belle \cite{Huschle:2015rga,Sato:2016svk,Hirose:2016wfn}, and LHCb \cite{Aaij:2015yra,Aaij:2017uff,Aaij:2017deq}. The averaging \emph{B}-tagged measurements of $R(D)$ and $R(D^{*})$ at the $\Upsilon(4 S)$ and the LHCb measurements of $R(D^{*})$ yield~\cite{Tanabashi:2018oca}
\begin{equation}
\begin{array}{l}{R(D)^{\mathrm{EX}}=0.407 \pm 0.039 \pm 0.024}, \\ {R\left(D^{*}\right)^{\mathrm{EX}}=0.304 \pm 0.013 \pm 0.007}.\end{array}
\end{equation}
Theoretically, there are already many precise SM predictions of these ratios. For example,
by fitting the lattice calculations and recent experimental data, Bigi and Gambino obtained \cite{Bigi:2016mdz}
\begin{equation}
R(D)^{\mathrm{SM}}=0.299 \pm 0.003.
\end{equation}
For $R(D^*)$, by using the heavy quark expansion and combining with the recent measurements of $\bar{B} \rightarrow D^{*} \ell \overline{\nu}_{\ell}$, Fajfer et al. obtained \cite{Fajfer:2012vx}
\begin{equation}
R\left(D^{*}\right)^{\mathrm{SM}}=0.252 \pm 0.003.
\end{equation}
Flavour Lattice Averaging Group (FLAG) combined recent lattice calculations and gave the average value \cite{Aoki:2016frl}
\begin{equation}
R(D)^{\mathrm{SM}}=0.300 \pm 0.008.
\end{equation}
We can easily see that the experimental values of $R(D)$ and $R(D^*)$ deviate from the SM predictions by $2.3\sigma$ and $3.4\sigma$ \cite{Tanabashi:2018oca}, respectively.

Most recently, LHCb reported the ratio of branching fractions \cite{Aaij:2017tyk}
\begin{equation}
R(J / \psi)^{\mathrm{EX}}=\frac{B r\left(B_{c} \rightarrow J / \psi \tau \nu\right)}{B r\left(B_{c} \rightarrow J / \psi \mu \nu\right)}=0.71 \pm 0.17 \pm 0.18.
\end{equation}
The SM predictions lie in the range $R(J/\psi)\in[0.23,0.29]$ , from which the data deviate by $2\sigma$.  To account for this deviation, both the new physics scenarios and the systematic errors were considered~\cite{Watanabe:2017mip,Tran:2018kuv,Bhattacharya:2018kig}.

The deviations of $R(D^{(*)})$ and $R(J/\psi)$ have motivated lots of theoretical studies on the semi-leptonic decays of $B_{(s)}$ to \emph{S}-wave charmed mesons. Besides papers mentioned above, the $B\rightarrow D(D^*)$ decays have been studied by QCD sum rules \cite{Bigi:2017jbd,Azizi:2008tt,Azizi:2008vt}, constituent quark models \cite{Polosa:2000ym}, Lattice QCD in the framework of heavy quark effective theory (HQET) \cite{Na:2015kha,Harrison:2017fmw}, and HQET method with the $O\left(\alpha_{s}, \Lambda_{\mathrm{QCD}} / m_{b, c}\right)$ and (part of) the $O(\Lambda_{\mathrm{QCD}}^{2} / m_{c}^{2})$ corrections \cite{Jung:2018lfu}, etc.

For the $B_{c} \rightarrow J / \psi\left(\eta_{c}\right)$ transitions, many other approaches, such as perturbative QCD (PQCD) \cite{Wen-Fei:2013uea}, QCD sum rules (QCDSR) \cite{Kiselev:2002vz}, light-cone QCD sum rules (LCSR) \cite{Fu:2018vap,Zhong:2018exo}, nonrelativistic QCD (NRQCD) \cite{Zhu:2017lqu,Shen:2014msa}, the covariant light-front quark model (CLFQM) \cite{Wang:2008xt}, the nonrelativistic quark model (NRQM) \cite{Hernandez:2006gt}, the relativistic quark model model (RQM) \cite{Ebert:2003cn}, the covariant confined quark model (CCQM) \cite{Tran:2018kuv} etc, have been used.

To explain the deviations, a lot of new physics models~\cite{Nierste:2008qe,Datta:2012qk,Becirevic:2012jf,Fajfer:2012jt,Crivellin:2012ye,Bauer:2015knc,Dorsner:2016wpm,Celis:2016azn} have been proposed. However, to make a reliable prediction of the NP, one needs both detections and theoretical calculations within the SM with more precision to study these observables $R(D^{(*)})$ and $R(J/\psi)$. For example, recently, the Belle collaboration presented an updated measurement of $R(D)$ which is $0.307\pm0.037\pm0.016$ \cite{Abdesselam:2019dgh}. It is in agreement with the SM prediction within $0.2\sigma$.

In this work, we will give a relativistic study of $R(D^{(*)})$ and $R(J/\psi)$ by using the improved instantaneous Bethe-Salpeter (BS) method. One of the essential parts of this method is the instantaneous BS wave function (also called Salpeter wave function) of mesons, which is achieved by solving the instantaneous BS equation (also called Salpeter equation). These functions are applied to calculate the hadronic transition matrix element. In our previous work~\cite{Wang:2012pf}, a similar method is used to study the channel $B\rightarrow D(D^*)$, where the results are not quite consistent with the experimental values. One possible reason is that we made approximations when boosting the wave functions of the final mesons to the initial meson rest frame. This method is improved in our another work \cite{Fu:2011tn} to study the rare decays of $B_c$ meson. Here we will systematically use this improved BS method to calculate the semi-leptonic decays of $B_q$ and $B_c$ mesons, and make more reliable predictions of $R(D^{(*)})$ and $R(J/\psi)$. Besides that, we will also give other quantities, including form factors,  $\mathcal{G}(1)$, the slope $\rho^{2}$, differential decay rate, branching ratios, etc.

The paper is organized as follows. In Section 2, we present the definitions of form factors of different decay channels and the differential decay width. In Section 3, we use the improved BS method to calculate the form factors. In Section 4, we give the numerical results, including the form factors, differential decay rate, partial decay widths, and the ratio of branching fractions. A conclusion is given finally.

\section{Formalism of semi-leptonic decays}

\begin{figure}[tbp]
	\centering
	
	\includegraphics[width=1.25\textwidth,trim=50 400 0 200,clip]{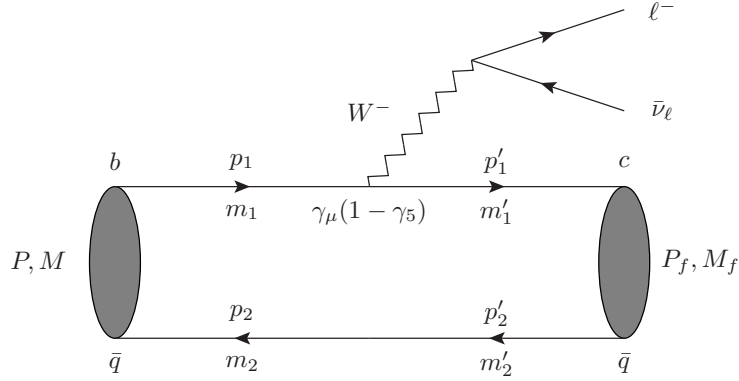}
	
	\caption{\label{fig:b-semi} Feynman diagram of the semileptonic $B_q$ decays  to a charmed $D_q^{(*)}$ ($q=u,d,s$).}
\end{figure}

In this section, we will present the formula of a $B_q$ ($q=u,d,s,c$) meson semi-leptonic decays to a charmed meson with the improved BS method. Fig.\ref{fig:b-semi} is the Feynman diagram of the semileptonic decay $B\rightarrow D_q l\bar\nu$, whose amplitude is written as
\begin{equation}
\label{e1}
\begin{split}
T=\frac{G_F}{\sqrt{2}} V_{bc}\bar{u}_{\ell}\gamma^\mu(1-\gamma_5)v_{\bar{\nu}_\ell} \langle D_q|J_\mu|B_q\rangle,
\end{split}
\end{equation}
where $G_F$ is the Fermi coupling constant, $V_{bc}$ is the CKM matrix element, $J_\mu=V_\mu-A_\mu$ is the charged eletroweak current.

The hadronic matrix element $\langle D_q^{(*)}|(V_\mu-A_\mu)|B_q^-\rangle$ can be characterized by the corresponding form factors.
If the final meson is a pseudoscalar state, the matrix element can be written as
\begin{equation}
\label{e2}
\begin{split}
\langle D_q|V^\mu|B_q\rangle &=F_+(Q^2)(P^\mu+P_f^\mu)+F_-(Q^2)(P^\mu-P_f^\mu)
\\
&=F_+(Q^2)(P^\mu+P_f^\mu)+F_-(Q^2)  Q^\mu
\\
 &=f_1(Q^2)(P^\mu+P^{\mu}_f-\frac{M^2-M^2_f}{Q^2}Q^\mu)+f_0(Q^2)\frac{M^2-M^2_f}{Q^2}Q^\mu,
\\
\langle D_q|A_\mu|B_q\rangle & =0,
\end{split}
\end{equation}
where $P$ and $P_f$ are the momenta of the initial and final mesons with masses $M$ and $M_f$, respectively; the definition $Q^\mu=P^\mu-P_f^\mu$ is used; $f_1(Q^2)$, $f_0(Q^2)$ are the form factors which are related to the functions $F_+(Q^2)$ and $F_-(Q^2)$ by
\begin{equation}
f_1(Q^2)=F_+(Q^2),\quad f_0(Q^2)=F_+(Q^2) + \frac{Q^2}{M^2-M_f^2}F_-(Q^2).
\end{equation}

If the final meson is a vector state, the matrix element is written as
\begin{equation}
\label{e25}
\begin{split}
\langle D^*_q|V_\mu|B_q\rangle
&=ig(Q^2) \varepsilon^{\mu \nu \alpha \tau}\epsilon_\nu P_{f\alpha} P_\tau
=\frac{2iV(Q^2)}{M+M_f} \varepsilon^{\mu \nu \alpha \tau}\epsilon_\nu P_{f\alpha} P_\tau,
\\
\langle D^*_q|A_\mu|B_q\rangle
& =f(Q^2)\epsilon^\mu+a_+(Q^2)(\epsilon\cdot q)(P^\mu+P^\mu)+a_-(Q^2)(\epsilon\cdot q)Q^\mu
\\
& =2M_f A_0(Q^2)\frac{\epsilon\cdot q}{Q^2}Q^\mu+(M+M_f)A_1(Q^2)[\epsilon^\mu-\frac{\epsilon\cdot q}{Q^2}Q^\mu] \\
&\quad -A_2(Q^2)\frac{\epsilon \cdot q}{M+M_f}[P^\mu+P_f^\mu-\frac{M^2-M_f^2}{Q^2}Q^\mu],
\end{split}
\end{equation}
where $\epsilon^\mu$ is the polarization vector of final meson $D^*_q$;   $\varepsilon^{\mu \nu \alpha \tau}$ is the totally antisymmetric Levi-Civita tensor;  $V(Q^2)$, $A_0(Q^2)$, $A_1(Q^2)$, $A_2(Q^2)$ are the form factors which are related to the functions $f(Q^2)$, $a_+(Q^2)$, $a_-(Q^2)$, $v(Q^2)$ by
\begin{equation}
\begin{split}
& V(Q^2)=\frac{M+M_f}{2}g(Q^2), \quad A_1(Q^2)=-\frac{f(Q^2)}{M+M_f}, \quad A_2(Q^2)=(M+M_f)a_+(Q^2)
\\
& A_0(Q^2)=-\frac{1}{2M_f}(Q^2 a_-(Q^2)+f(Q^2)+(M^2-M_f^2) a_+(Q^2)).
\end{split}
\end{equation}

The square of the transition amplitude can be written as
\begin{equation}
\label{e4}
\begin{split}
\sum |T|^2=\frac{G_F^2}{2}|V_{bc}|^2 L_{\mu \nu}H^{\mu \nu},
\end{split}
\end{equation}
where we have summed up the possible polarization of finial state.  $L_{\mu \nu}$ is the leptonic tensor, which has the form
\begin{equation}
\label{e5}
\begin{split}
L_{\mu \nu} &
\equiv \overline{u}_{\ell} \gamma_{\mu}\left(1-\gamma_{5}\right) v_{\bar\nu_\ell} \overline{v}_{\bar\nu_\ell}\left(1+\gamma_{5}\right) \gamma_{\nu} u_{\ell}
\\
& = 8\left(P_{l \mu} P_{\bar{\nu} \nu}+P_{\bar{\nu} \mu} P_{l \nu}-g_{\mu \nu} P_{l} \cdot P_{\bar{\nu}}-i \epsilon_{\mu \nu \rho \sigma} P_{l}^{\rho} P_{\bar{\nu}}^{\sigma}\right),
\end{split}
\end{equation}
where $P_{l}$ and $P_{\bar\nu}$ are the momenta of $l$ and $\bar\nu$, respectively.
The hadronic tensor $H^{\mu\nu}$ can be written as
\begin{equation}
\label{e6}
\begin{split}
H^{\mu \nu} & \equiv \sum\langle D_q|J^\mu |B_q\rangle\langle B_q|J^\nu|D_q\rangle
\\
& =-\alpha g^{\mu \nu}
\\
&\quad +   \beta_{++}(P+P_f)^\mu(P+P_f)^\nu  +  \beta_{+-}(P+P_f)^\mu(P-P_f)^\nu
\\
&\quad + \beta_{-+}(P-P_f)^\mu(P+P_f)^\nu  +  \beta_{--}(P-P_f)^\mu(P-P_f)^\nu
\\
&\quad +  i\gamma\epsilon^{\mu \nu \rho \sigma}(P+P_f)_\rho (P-P_f)_\sigma,
\end{split}
\end{equation}
where the functions $\alpha$, $\beta_{++}$, $\beta_{+-}$, $\beta_{-+}$, $\beta_{--}$ and $\gamma$ directly relate to the form factors. For the decays when the final state is a $0^-$ meson, we have
\begin{equation}
\label{e71}
\begin{split}
\alpha=&\gamma=0,
\\
\beta_{++}=F_+^2, \quad &\beta_{--}=F_-^2,
\\
\beta_{+-}=F_+F_-, \quad &\beta_{-+}=F_-F_+.
\end{split}
\end{equation}
When the final state is a $1^-$ meson, the relations are
\begin{equation}
\label{eb1}
\begin{split}
\alpha &= f^2+4M^2g^2|\vec{P}_f|^2,
\\
\beta_{++} &= \frac{f^2}{4M_f^2}-M^2g^2y+\frac{1}{2}\left[ \frac{M^2}{M_f^2}(1-y)-1  \right]fa_++\frac{M^2|\vec{P}_f|^2}{M_f^2}a_+^2,
\\
\beta_{+-} &= \beta_{-+}= g^2(M^2-M_f^2)-\frac{f^2}{4M_f^2}-\frac{1}{2}f(a_++a_-)-\frac{1}{2}(a_+-a_-)\frac{ME_f}{M_f^2}+a_+a_-\frac{M^2|\vec{P}_f|^2}{M_f^2},
\\
\beta_{--} &= -g^2(M^2+2ME_f+M_f^2)+\frac{f^2}{4M_f^2}-fa_-\left( \frac{ME_f}{M_f^2}  \right)+a^2_-\frac{M^2|\vec{P}_f|^2}{M_f^2},
\\
\gamma &= 2gf.
\end{split}
\end{equation}

Finally, the decay width $\Gamma$ is read as
\begin{equation}
\label{e7}
\begin{split}
\Gamma=\frac{1}{2M(2\pi)^9} \int{\frac{d^3 \vec{P_f}}{2E_f} \frac{d^3\vec{P_\ell}}{2E_\ell} \frac{d^3\vec{P}_{\bar\nu}}{2E_{\bar\nu}} (2\pi)^4 \delta^4(P-P_f-P_\ell-P_{\bar\nu}) \sum |T|^2},
\end{split}
\end{equation}
where $E_f$, $E_\ell$ and $E_{\bar\nu}$ the energies of $D^{(*)}_q$, $\ell$ and $\bar{\nu}_\ell$, respectively. By introducing the symbols $
x\equiv E_\ell/M$, $\quad y\equiv (P-P_f)^2/M^2$, the differential decay width can be written as
\begin{equation}
\label{e8}
\begin{split}
\frac{d^2\Gamma}{dxdy} =& |V_{cs}|^2 \frac{G_F^2 M^5}{64\pi^3}
\left\{ \frac{2\alpha}{M^2} \left( y-\frac{m^2_\ell}{M^2} \right)  \right.
\\
& +\beta_{++}\left[ 4\left(  2x \left( 1-\frac{M^2_f}{M^2}+y \right) -4x^2-y \right) + \frac{m^2_\ell}{M^2}\left( 8x+4\frac{M^2_f}{M^2}-3y-\frac{m^2_\ell}{M^2} \right) \right]
\\
&+\left(  \beta_{+-}+\beta_{-+} \right) \frac{m^2_\ell}{M^2} \left( 2-4x+y-2\frac{M^2_f}{M^2}+\frac{m^2_\ell}{M^2} + \beta_{--}\frac{m^2_\ell}{M^2}\left( y-\frac{m^2_\ell}{M^2} \right) \right)
\\
&\left. -\left[  2\gamma y\left( 1-\frac{M^2_f}{M^2}-4x+y+\frac{m^2_\ell}{M^2} \right) +2\gamma\frac{m^2_\ell}{M^2} \left( 1-\frac{M^2_f}{M^2} \right) \right] \right\}.
\end{split}
\end{equation}
And the decay width is
\begin{equation}
\Gamma=\int{dx \int{dy \frac{d^2\Gamma}{dxdy}}}.
\end{equation}

\section{The improved BS method}

The matrix element $\langle D_q^{(*)}|J_\mu|B_q^-\rangle$ will be calculated by the improved BS method. Within Mandelstam formalism, it can be written as
\begin{equation}\label{eq:amporigin}
\begin{aligned}
\left\langle P_{f}\left|J^{\mu}\right| P\right\rangle&=\int \frac{d^{4} q}{(2 \pi)^{4}} \frac{d^{4} q_{f}}{(2 \pi)^{4}} \operatorname{Tr}\left[\overline{\chi}(P_{f},q_{f}) \Gamma^{\mu} \chi({P},q) S_{2}^{-1}\left(-p_{2}\right)\right](2 \pi)^{4} \delta^{4}\left(p_{2}-p_{2f}\right)\\
&=\left.\int \frac{d^{4} q}{(2 \pi)^{4}} \operatorname{Tr}\left[\overline{\chi}\left(P_{f}, q_{f}\right) \Gamma^{\mu} \chi(P, q) S_{2}^{-1}\left(-p_{2}\right)\right]\right|_{q_{f}=q+\alpha_{2 f} P_{f}-\alpha_{2} P}
\end{aligned}
\end{equation}
where $\chi({P},q)$ and $\bar{\chi}({P},q)$ are the BS wave function of the initial meson and final meson, respectively, and the latter one has the form $\overline{\chi}(P_{f},q_{f})=\gamma_0 \chi(P_{f},q_{f})^{\dag}\gamma_0$ in its rest frame; the vertex is $\Gamma^\mu=\gamma^\mu(1-\gamma^5)$; $S_1$ and $S_2$ are propagators of the quark and anti-quark, respectively. $q$ and is the relative momentum of the quark and antiquark within the initial meson. $p_1$, $p_2$ are respectively the momenta of the quark and anti-quark within the initial meson, which are related to $P$ and $q$ by
\begin{equation}
p_i=\alpha_iP+Jq,\qquad \alpha_i\equiv\frac{m_i}{m_1+m_2},
\end{equation}
where $m_1$, $m_2$ are the masses of the quark and anti-quark, respectively; $J=1$ and $-1$ for the cases $i=1$ and $2$, respectively. For the final meson, we define similar relations
\begin{equation}
p_{if}=\alpha_{if}P_f+Jq_f,\qquad \alpha_{if}\equiv\frac{m_{if}}{m_{1f}+m_{2f}}.
\end{equation}

The BS wave functions fulfill the BS equation which has the form \cite{Chang:2006tc}
\begin{equation}
\label{ea1}
\begin{split}
({p\!\!\!/}_1-m_1)\chi(P,q)({p\!\!\!/}_2+m_2)=i\int{\frac{d^4k}{(2\pi)^4}V(P,k,q)\chi(P,k)},
\end{split}
\end{equation}
where $V(P,k,q)$ is the interaction kernel. If we take the instantaneous approximation, the kernel can be reduced to $V(|\vec q-\vec k|)$. Now we can introduce two 3-dimensional quantities
\begin{equation}
\begin{split}
\varphi(q_\perp) \equiv i \int{\frac{dq_{_P}}{2\pi} \chi(P,q)},
\qquad
\eta(P,q_\perp) \equiv \int{\frac{dk^3_\perp}{(2\pi)^3} V(k_\perp,q_\perp)\varphi(k_\perp) }.
\end{split}
\end{equation}
where we have used the definitions
\begin{equation}
q_{_P}=\frac{P\cdot q}{M},~~~~~ q^\mu_\perp= q^\mu-q_PP^\mu.
\end{equation}
Then the BS equation can be rewritten as
\begin{equation}
\label{ea3}
\begin{aligned}
\chi(P,q)=S_1(p_1)\eta(P,q_\perp)S_2(-p_2).
\end{aligned}
\end{equation}

Using above equations, we can write Eq.~\eqref{eq:amporigin} as
\begin{equation}\label{aee1}
\begin{aligned}
\left\langle P_{f}\left|J^{\mu}\right| P\right\rangle
&=\int\frac{d^4q}{(2\pi)^4}\operatorname{Tr}\left[\overline\eta(P_f, q_{f\ci})S_1(p_{1f})\Gamma^\mu S_1(p_1)\eta(P, q_\perp)S_2(-p_{2})\right] \\
&=\int\frac{d^4q}{(2\pi)^4}{\rm Tr}\Big[\frac{\slashed P_f}{M_f}(\widetilde{\Lambda}^+_1(p_{1f\ci})+\widetilde{\Lambda}^-_1(p_{1f\ci}))\overline\eta(P_f, q_{f\ci})(\widetilde{\Lambda}^+_2(p_{2f\ci})\\
&+\widetilde{\Lambda}^-_2(p_{2f\ci}))\frac{\slashed P_f}{M_f}S_1(p_{1f})\Gamma^\mu S_1(p_1)\eta(P, q_\perp)S_2(-p_{2})\Big]\\
&\approx\int\frac{d^4q}{(2\pi)^4}{\rm Tr}\Big[\frac{\slashed P_f}{M_f}\widetilde{\Lambda}^+_1(p_{1f\ci})\overline\eta(P_f, q_{f\ci})\widetilde{\Lambda}^+_2(p_{2f\ci})\frac{\slashed P_f}{M_f}\\
&\times S_1(p_{1f})\Gamma^\mu S_1(p_1)\eta(P, q_\perp)S_2(-p_{2})\Big].
\end{aligned}
\end{equation}
In the first line of the above equation, we have used Eq.~\eqref{ea3} with the definition $q_{f\ci}\equiv q_f-\frac{P_f\cdot q_f}{M_f^2}P_f$ for the final meson; in the second line, we have defined the projection operator of the final meson
\begin{equation}
\widetilde\Lambda^\pm_i(p_{if\ci})=\frac{1}{2\widetilde\omega_{if}}\Big[\frac{\slashed P_f}{M_f}\widetilde\omega_{if} \pm(Jm_{if}+\slashed p_{if\ci})\Big],
\end{equation}
with
\begin{equation}
\begin{aligned}
\widetilde\omega_{if}\equiv\sqrt{m_{if}^2-p_{if\ci}^2}=\sqrt{m_{if}^2-q_{f\ci}^2}.
\end{aligned}
\end{equation}
The relation $\frac{\slashed P_f}{M_f}=\widetilde{\Lambda}^+_1(p_{1f\ci})+\widetilde{\Lambda}^-_1(p_{1f\ci})$ is also applied. In the last equation, we have omitted the contribution of the negative energy part, which is very small compared with that of the positive energy part.

Next, we express the propagators $S_i(Jp_i)$ and $S_i(Jp_{if})$ also in terms of the projection operators,
\begin{equation}
\begin{aligned}
&S_i(Jp_{i}) = \frac{\Lambda^+_i(p_{i\perp})}{p_{i_P}-\omega_{i}+i\epsilon} + \frac{\Lambda^-_i(p_{i\perp})}{p_{i_P}+\omega_{i}-i\epsilon},\\
&S_i(p_{if}) = \frac{\Lambda^+_i(p_{if\perp})}{p_{if_P}-\omega_{if}+i\epsilon} + \frac{\Lambda^-_i(p_{if\perp})}{p_{if_P}+\omega_{if}-i\epsilon},
\end{aligned}
\end{equation}
where
\begin{equation}
\begin{aligned}
&\Lambda^\pm_i(p_{i\perp})=\frac{1}{2\omega_{i}}\Big[\frac{\slashed P}{M}\omega_{i} \pm(Jm_{i}+\slashed p_{i\perp})\Big],~~~
\omega_{i}\equiv\sqrt{m_{i}^2-p_{i\perp}^2}=\sqrt{m_{i}^2-q_{\perp}^2},\\
&\Lambda^\pm_i(p_{if\perp})=\frac{1}{2\omega_{if}}\Big[\frac{\slashed P}{M}\omega_{if} \pm(Jm_{i}+\slashed p_{if\perp})\Big],~~~
\omega_{if}\equiv\sqrt{m_{if}^2-p_{if\perp}^2}.\\
\end{aligned}
\end{equation}
Then Eq.~\eqref{aee1} can be written as
\begin{equation}
\begin{aligned}
\langle P_f|J^{\mu}|P\rangle&=\int\frac{d^4q}{(2\pi)^4}{\rm Tr}\Big[\frac{\slashed P_f}{M_f}\widetilde{\Lambda}^+_1(p_{1f\ci})\overline\eta(P_f, q_{f\ci})\widetilde{\Lambda}^+_2(p_{2f\ci})\frac{\slashed P_f}{M_f}\Lambda^+_1(p_{1f\perp})\Gamma^\mu \Lambda^+_1(p_{1\perp})\\
&~\times \eta(P, q_\perp)\Lambda^+_2(p_{2\perp})\Big]\frac{1}{(p_{1f_P}-\omega_{1f}+i\epsilon)(p_{1_P}-\omega_{1}+i\epsilon)(p_{2_P}-\omega_{2}+i\epsilon)},
\end{aligned}
\end{equation}
where the quantities $p_{1_P}$, $p_{2_P}$, and $p_{1f_P}$ in the denominator are related to $q_{_P}$ by
\begin{equation}
\begin{aligned}
&p_{1_P}  = q_{_P}+\alpha_1M,\\
&p_{2_P} = -q_{_P} +\alpha_2 M,\\
&p_{1f_P}  = q_{_P} +P_{f_P} -\alpha_2 M.
\end{aligned}
\end{equation}
By integrating out $q_{_P}$ around the upper plane, we get
\begin{equation}
\begin{aligned}
\langle P_f|J^{\mu}|P\rangle&=-i\int\frac{d\vec q}{(2\pi)^3}{\rm Tr}\Big[\frac{\slashed P_f}{M_f}\widetilde{\Lambda}^+_1(p_{1f\ci})\overline\eta(P_f, q_{f\ci})\widetilde{\Lambda}^+_2(p_{2f\ci})\frac{\slashed P_f}{M_f}\Lambda^+_1(p_{1f\perp})\Gamma^\mu\\
&~\times  \Lambda^+_1(p_{1\perp})\eta(P, q_\perp)\Lambda^+_2(p_{2\perp})\Big]\frac{1}{(P_{f_P}-\omega_{2} -\omega_{1f})}\frac{1}{(M-\omega_{2} -\omega_{1})}.
\end{aligned}
\end{equation}

The 3-dimensional wave functions (Salpeter wave function) of the initial and final mesons fulfill corresponding Salpeter equations
\begin{equation}
\begin{aligned}
&\varphi^{++}(P,q_\perp) =\frac{\Lambda^+_1(p_{1\perp}) \eta(P,q_\perp ) \Lambda^+_2(p_{2\perp})}{M-\omega_1-\omega_2},
\\
&\varphi^{++}(P_f,q_{f\ci}) =\frac{\widetilde\Lambda^+_1(p_{1f\ci}) \eta(P_f,q_{f\ci}) \widetilde\Lambda^+_2(p_{2f\ci})}{M_f-\widetilde\omega_{1f}-\widetilde\omega_{2f}},
\\
\end{aligned}
\end{equation}
where we have used the definitions
\begin{equation}
\begin{aligned}
&\varphi^{++}(P,q_\perp) = \Lambda_1^{+}(p_{1\perp})\frac{\slashed P}{M}\varphi(P, q_\perp)\frac{\slashed P}{M}\Lambda_2^{+}(p_{2\perp}),\\
&\varphi^{++}(P_f,q_{f\ci}) = \widetilde\Lambda_1^{+}(p_{1f\ci})\frac{\slashed P_f}{M_f}\varphi(P_f, q_{f\ci})\frac{\slashed P_f}{M_f}\widetilde\Lambda_2^{+}(p_{2f\ci}),\\
\end{aligned}
\end{equation}
whose explicit form can be found in Eq.\eqref{Eq:0-} and Eq.\eqref{Eq:1-}. Then the hadronic transition matrix element is written as \cite{Fu:2011tn}
\begin{equation}
\begin{aligned}
\left\langle P_{f}\left|J^{\mu}\right|P\right\rangle&=-i\int\frac{d\vec q}{(2\pi)^3}{\rm Tr}\Big[\frac{\slashed P_f}{M_f}\overline\varphi^{++}(P_f, q_{f\ci})\frac{\slashed P_f}{M_f}L_{r}\Gamma^\mu \varphi^{++}(P, q_\perp)\Big],
\end{aligned}
\end{equation}
where
\begin{equation}
L_{r}=\frac{\left(M_{f}-\widetilde\omega_{1f}-\widetilde\omega_{2f}\right)}{\left(P_{f_{P}}-\omega_{1 f}-\omega_2\right)} \Lambda_{1}^{+}\left(p_{1f\perp}\right).
\end{equation}

Here we use the relativistic wave function for a $0^-$ meson, which has the form
\begin{equation}\label{Eq:0-}
\varphi_{0^{-}}\left(q_{\perp}\right)=M\left[\frac{P\!\!\!/}{M} \varphi_{1}\left(q_{\perp}\right)+\varphi_{2}\left(q_{\perp}\right)+\frac{q\!\!\!/_{\perp}}{M} \varphi_{3}\left(q_{\perp}\right)+\frac{P\!\!\!/ q\!\!\!/_{\perp}}{M^{2}} \varphi_{4}\left(q_{\perp}\right)\right] \gamma_{5},
\end{equation}
where the radial wave functions $\varphi_1\sim\varphi_4$ fulfill the constraint conditions
\begin{equation}
\begin{aligned} \varphi_{3} &=\frac{M\left(\omega_{2}-\omega_{1}\right)}{m_{1} \omega_{2}+m_{2} \omega_{1}} \varphi_{2}, \\ \varphi_{4} &=-\frac{M\left(\omega_{1}+\omega_{2}\right)}{m_{1} \omega_{2}+m_{2} \omega_{1}} \varphi_{1}.
\end{aligned}
\end{equation}
The numerical values of $\varphi_{1}$ and $\varphi_{2}$ can be obtained by solving the Salpeter equation.

For the $1^-$ state, the relativistic wave function has the form
\begin{equation}\label{Eq:1-}
\begin{aligned}
\varphi_{1^{-}}\left(q_{\perp}\right)  = & \left(q_{\perp} \cdot \epsilon\right)\left[f_{1}\left(q_{\perp}\right)+\frac{P\!\!\!/}{M} f_{2}\left(q_{\perp}\right)+\frac{q\!\!\!/_{\perp}}{M} f_{3}\left(q_{\perp}\right)+\frac{P\!\!\!/ q\!\!\!/_{\perp}}{M^{2}} f_{4}\left(q_{\perp}\right)\right],
\\
& +M \epsilon\!\!\!/ \left[f_{5}\left(q_{\perp}\right)+\frac{P\!\!\!/}{M} f_{6}\left(q_{\perp}\right)+\frac{q\!\!\!/_{\perp}}{M} f_{7}\left(q_{\perp}\right)+\frac{P\!\!\!/ q\!\!\!/_{\perp}}{M^{2}} f_{8}\left(q_{\perp}\right)\right].
\end{aligned}
\end{equation}
And the radial wave functions $f_1\sim f_8$ fulfill the constraint conditions
\begin{equation}
\begin{aligned} f_{1}\left(q_{\perp}\right) &=\frac{q_{\perp}^{2} f_{3}\left(\omega_{1}+\omega_{2}\right)+2 M^{2} f_{5} \omega_{2}}{M\left(m_{1} \omega_{2}+m_{2} \omega_{1}\right)}, \\ f_{2}\left(q_{\perp}\right) &=\frac{q_{\perp}^{2} f_{4}\left(\omega_{1}-\omega_{2}\right)+2 M^{2} f_{6} \omega_{2}}{M\left(m_{1} \omega_{2}+m_{2} \omega_{1}\right)}, \\ f_{7}\left(q_{\perp}\right) &=\frac{M\left(\omega_{1}-\omega_{2}\right)}{m_{1} \omega_{2}+m_{2} \omega_{1}} f_{5}, \\ f_{8}\left(q_{\perp}\right) &=\frac{M\left(\omega_{1}+\omega_{2}\right)}{m_{1} \omega_{2}+m_{2} \omega_{1}} f_{6}. \end{aligned}
\end{equation}

For comparison, we also present the non-relativistic forms of the wave functions, which have the form
\begin{equation}
\varphi_{0^{-}}\left(q_{\perp}\right)=M\left[\frac{P\!\!\!/}{M} \varphi_{1}\left(q_{\perp}\right)+\varphi_{2}\left(q_{\perp}\right)\right] \gamma_{5}
\end{equation}
and 
\begin{equation}
\begin{aligned}
\varphi_{1^{-}}\left(q_{\perp}\right)  = M \epsilon\!\!\!/ \left[f_{3}\left(q_{\perp}\right)+\frac{P\!\!\!/}{M} f_{4}\left(q_{\perp}\right)\right]
\end{aligned}
\end{equation}
for the $0^-$ and $1^-$ mesons, respectively.

\section{Numerical Results and Discussions}
In this work, we use the Cornell potential as the interaction kernel \cite{Kim:2003ny},  which is a linear scalar potential plus a vector interaction potential
\begin{equation}
\label{ea12}
\begin{split}
V(\vec{q}) & =V_s(\vec{q})+V_v(\vec{q})\gamma_0\otimes\gamma^0,
\\
V_s(\vec{q}) & =-\left( \frac{\lambda}{\alpha} +V_0 \right) \delta^3(\vec{q}) + \frac{\lambda}{\pi^2}\frac{1}{(\vec{q}^2+\alpha^2)^2},
\\
V_v(\vec{q}) & =-\frac{2}{3\pi^2}\frac{\alpha_s(\vec{q})}{(\vec{q}^2+\alpha^2)},
\end{split}
\end{equation}
where the QCD running coupling constant $\alpha_s(\vec{q})=\frac{12\pi}{33-2N_f}\frac{1}{\mathtt{log}\left( a+\vec{q}^2/\Lambda^2_{QCD} \right)}$ is used;  the symbol $\bigotimes$ denotes that the Salpeter wave function is sandwiched between the two $\gamma^0$ matrices. The constants $\lambda$, $\alpha$, $a$, $V_0$ and $\Lambda_{QCD}$ are the parameters charactering the potential, which have the values~\cite{Wang:2012cp}, 
\begin{equation}
\begin{array}{llr}{a=e=2.7183,} & {\alpha=0.060 ~\mathrm{GeV},} & {\lambda=0.210~ \mathrm{GeV}^{2}}, \\ {m_{u}=0.305 ~\mathrm{GeV},} & {m_{d}=0.311 ~\mathrm{GeV},} & {m_{s}=0.500~ \mathrm{GeV}}, \\ {m_{c}=1.62 ~\mathrm{GeV},} & {m_{b}=4.96~ \mathrm{GeV},} & {\Lambda_{\mathrm{QCD}}=0.270 ~ \mathrm{GeV}}.\end{array}
\end{equation}
In addition, the CKM matrix element $|V_{c b}|=0.0411$ from PDG \cite{Tanabashi:2018oca} is also used. 

\begin{figure}[h]
  \begin{minipage}[t]{0.5\textwidth}
   \centering
   \includegraphics[width=3.4in]{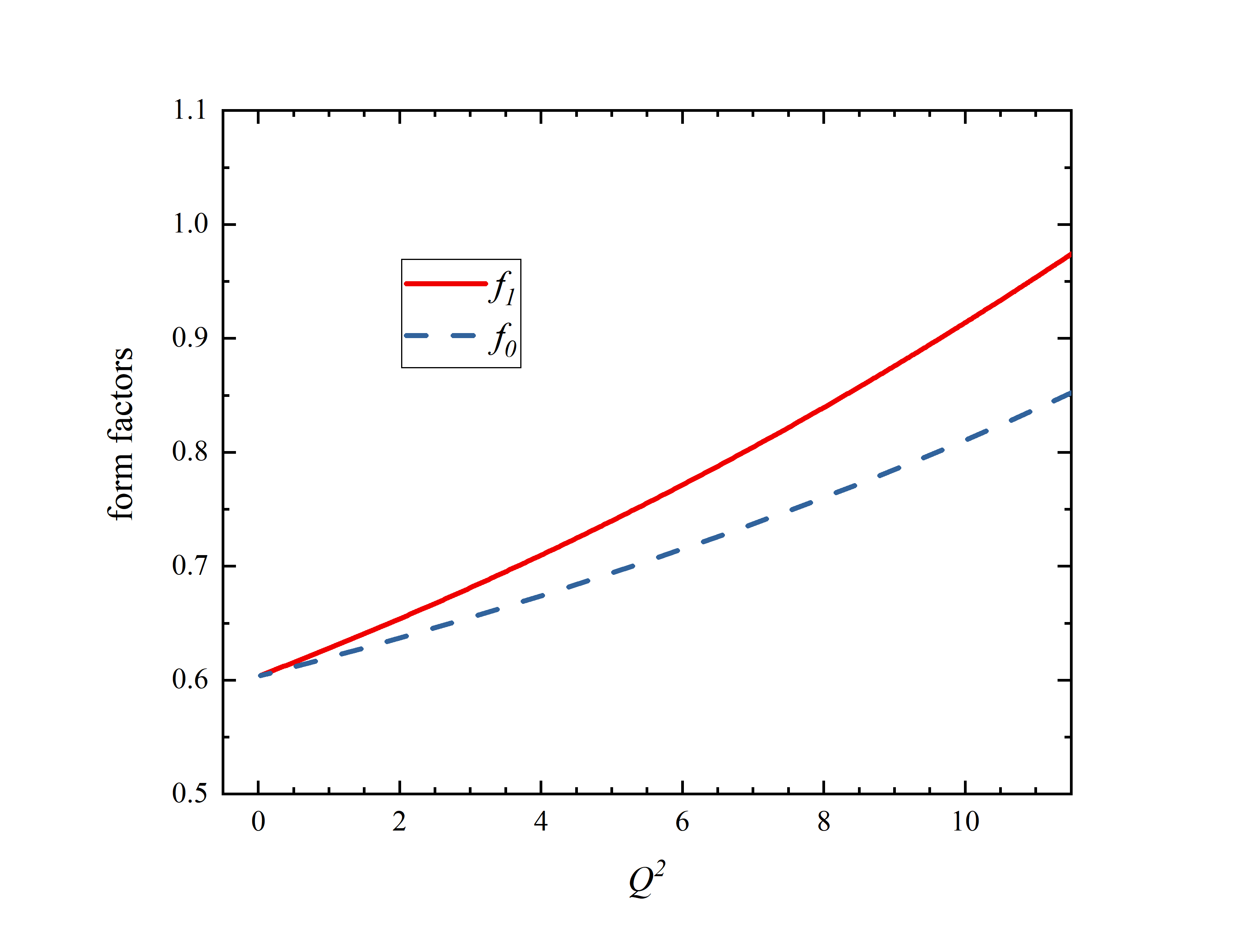}
   \subcaption{$f_1(Q^2)$, $f_0(Q^2)$ of $0^- \rightarrow 0^-$  }
   \label{fig:side:a}
  \end{minipage}%
  \begin{minipage}[t]{0.5\textwidth}
   \centering
   \includegraphics[width=3.4in]{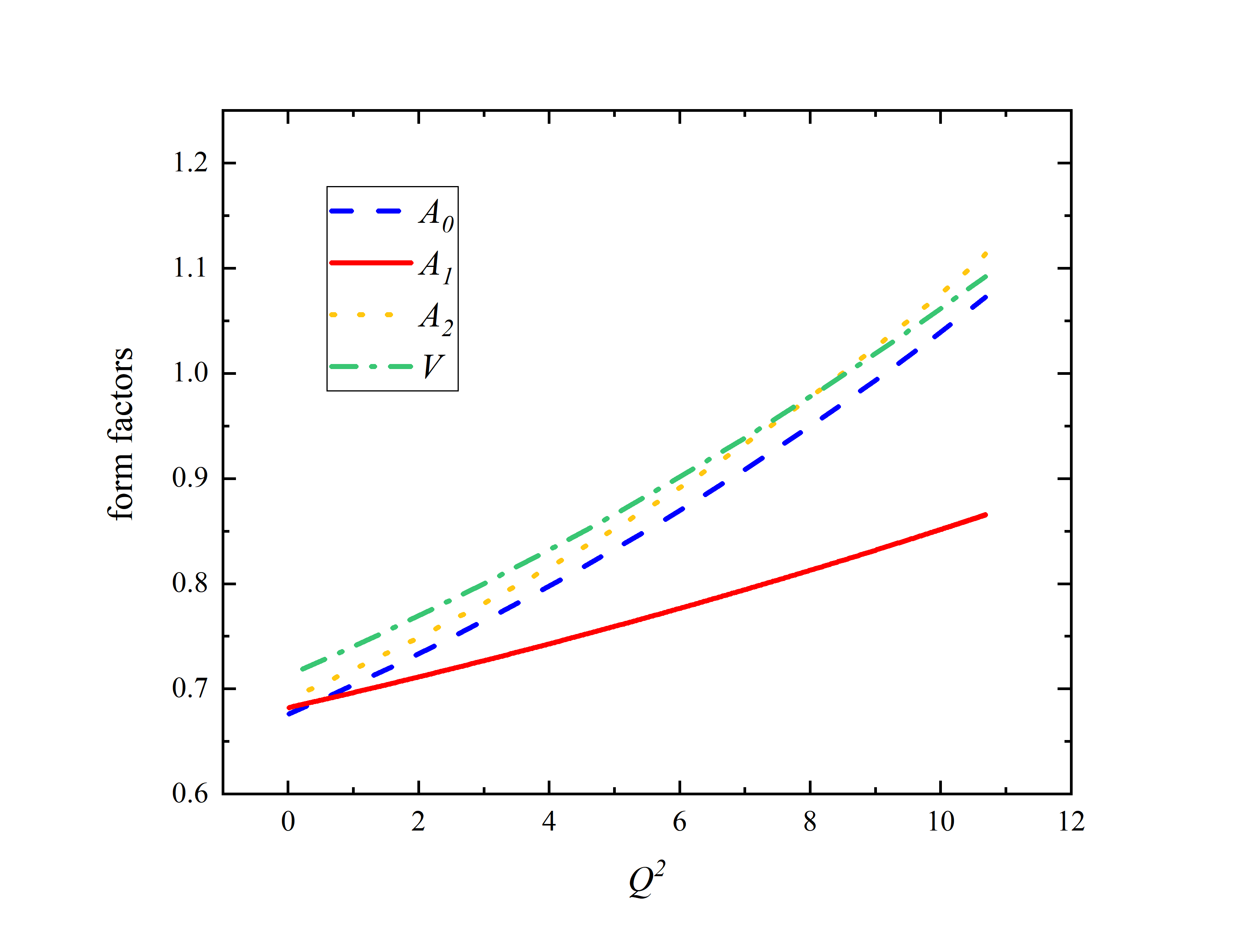}
   \subcaption{$A_0(Q^2)$, $A_1(Q^2)$, $A_2(Q^2)$, $V(Q^2)$ of $0^- \rightarrow 1^-$ }
   \label{fig:side:b}
  \end{minipage}
  \caption{\label{fig:a24} The form factors of decays $\bar{B}^0\rightarrow D^{(*)+}e \bar{\nu}_e$. }
\end{figure}

\begin{figure}[h]
  \begin{minipage}[t]{0.5\textwidth}
   \centering
   \includegraphics[width=3.4in]{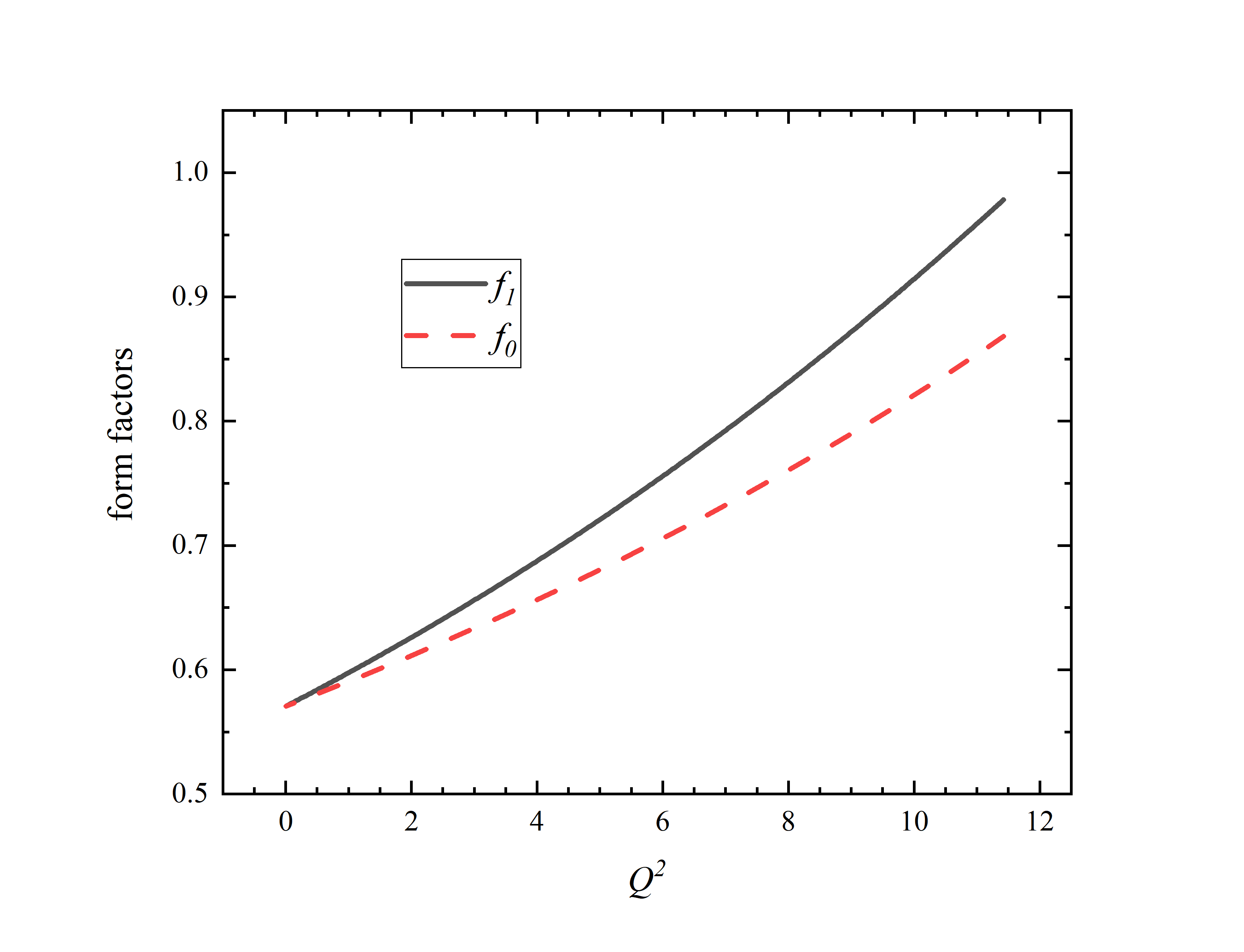}
   \subcaption{$f_1(Q^2)$, $f_0(Q^2)$ of $0^- \rightarrow 0^-$}
  \end{minipage}%
  \begin{minipage}[t]{0.5\textwidth}
   \centering
   \includegraphics[width=3.4in]{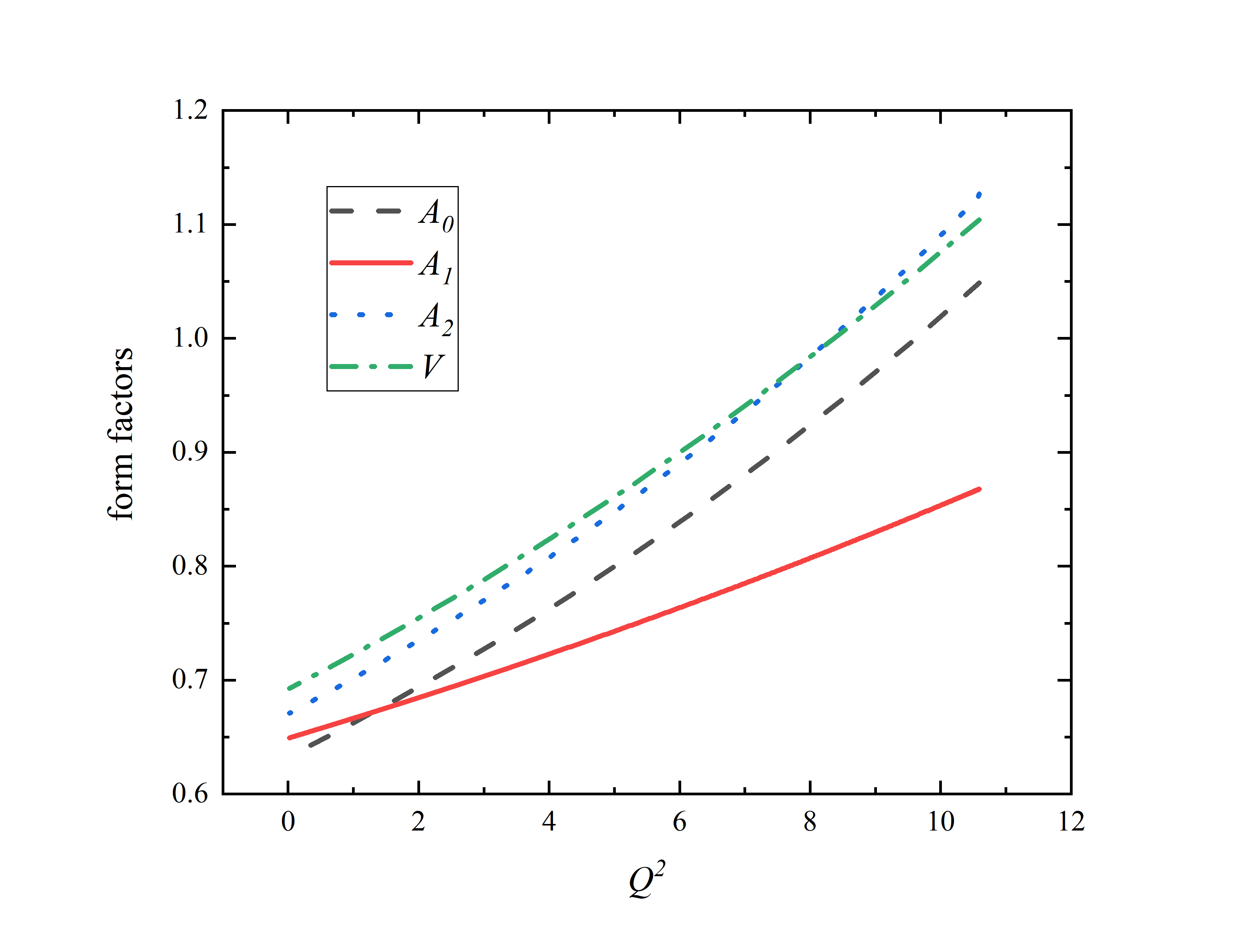}
   \subcaption{$A_0(Q^2)$, $A_1(Q^2)$, $A_2(Q^2)$, $V(Q^2)$ of $0^- \rightarrow 1^-$ }
  \end{minipage}
  \caption{\label{fig:a25} The form factors of decays $B_s^0\rightarrow D_s^{(*)+}e \bar{\nu}_e$. }
\end{figure}

\begin{figure}[h]
  \begin{minipage}[t]{0.5\textwidth}
   \centering
   \includegraphics[width=3.4in]{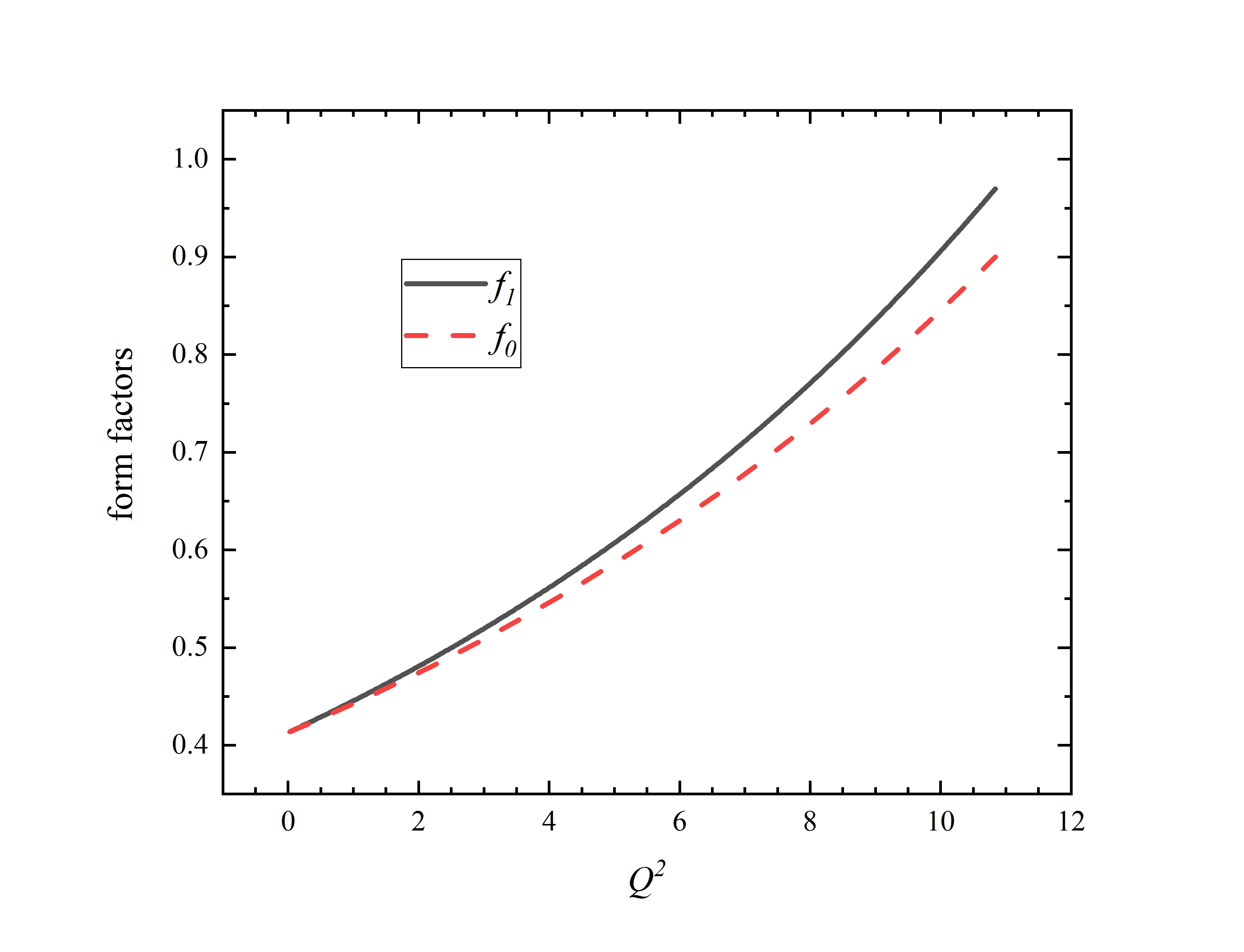}
   \subcaption{$f_1(Q^2)$, $f_0(Q^2)$ of $0^- \rightarrow 0^-$}
  \end{minipage}%
  \begin{minipage}[t]{0.5\textwidth}
   \centering
   \includegraphics[width=3.4in]{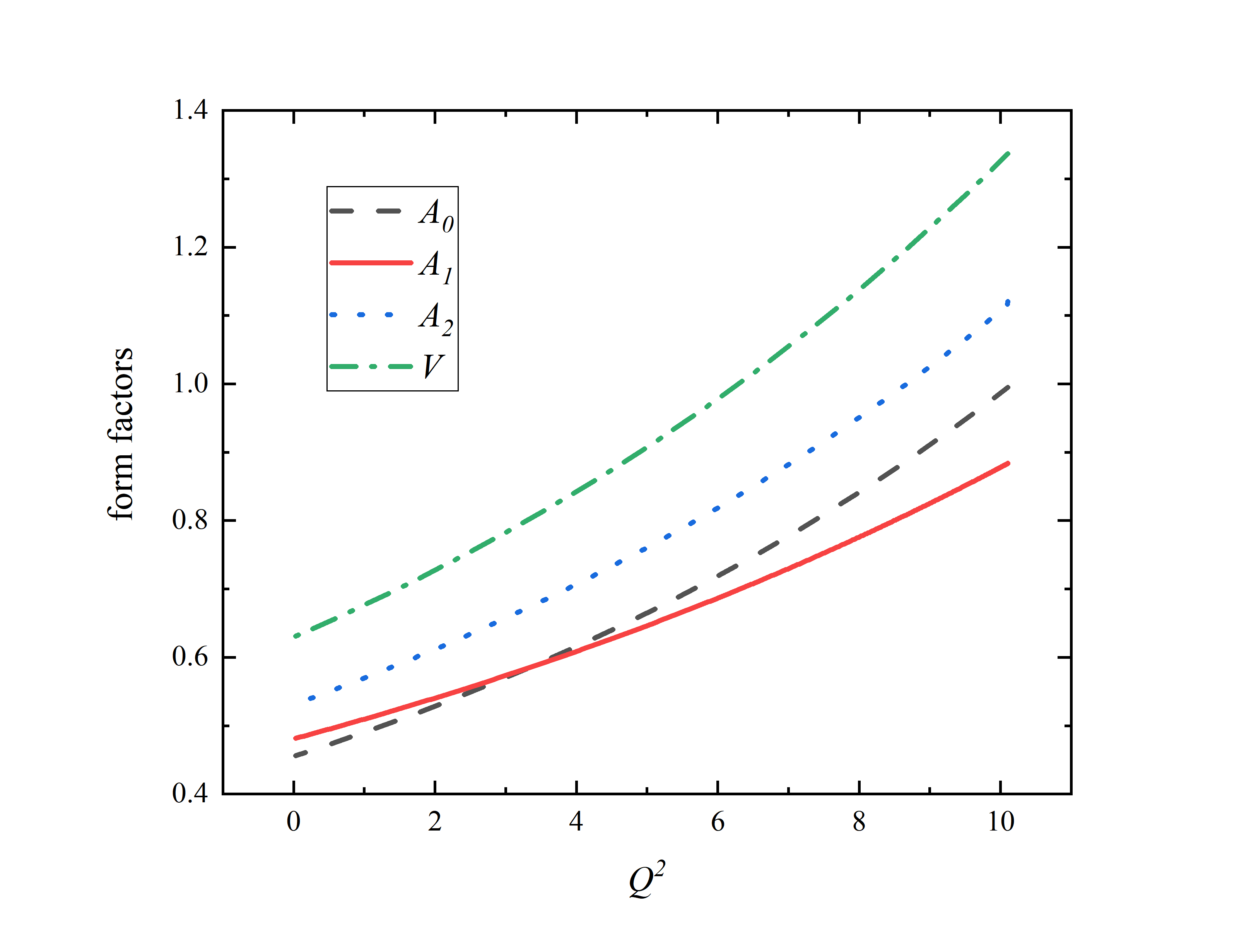}
   \subcaption{$A_0(Q^2)$, $A_1(Q^2)$, $A_2(Q^2)$, $V(Q^2)$ $0^- \rightarrow 1^-$ }
  \end{minipage}
  \caption{\label{fig:a26}The form factors of decays $B_c^-\rightarrow \eta_c(J/\psi)e \bar{\nu}_e$. }
\end{figure}

Since the Salpeter equations of the $0^-$ and $1^-$ mesons have been solved in our previous paper~\cite{Kim:2003ny,Wang:2005qx}, we will not show the details, but directly give the numerical results of wave functions. With Eq.\eqref{e2} and Eq.\eqref{e25}, we can get the form factors of $\bar{B}^0\rightarrow D^{+}e \nu_e$, $\bar{B}^{0}\rightarrow D^{*+}e \nu_e$ and $B_c\rightarrow \eta_c(J/\psi) e\nu_e$ which are presented in Fig.\ref{fig:a24}, Fig.\ref{fig:a25} and Fig.\ref{fig:a26}, respectively. In each figure, we plot two diagrams, the left one is for the case when the final state is a pseudoscalar, and the right one for the vector final state.

To check our result, we compare it with that achieved by other method, where the form factors are parameterized in a different way. According to Ref.~\cite{Amhis:2016xyh}, The differential width can be written as
\begin{equation}\label{eqbr}
\begin{split}
\frac{d \Gamma\left({B} \rightarrow D \ell \overline{\nu}_{\ell}\right)}{d w}=\frac{G_{\mathrm{F}}^{2} m_{D}^{3}}{48 \pi^{3}}\left(m_{B}+m_{D}\right)^{2}\left(w^{2}-1\right)^{3 / 2} \eta_{\mathrm{EW}}^{2} \mathcal{G}^{2}(w)\left|V_{c b}\right|^{2},
\end{split}
\end{equation}
where the factor $\eta_{\mathrm{EW}}=1+\alpha / \pi \ln M_{Z} / m_{B} \simeq 1.0066$ takes into account the short distance QED corrections. Moreover, the recoil variable $w$ is defined as the product of the 4-velocities of the $B$ and $D$ mesons, which is related to $Q^2$ by the formula
\begin{equation}\label{eqw}
w=v_{B} \cdot v_{D}=\frac{m_{B}^{2}+m_{D}^{2}-Q^{2}}{2 m_{B} m_{D}}.
\end{equation}

\begin{figure}[h]
  \begin{minipage}[t]{0.5\textwidth}
   \centering
   \includegraphics[width=3.4in]{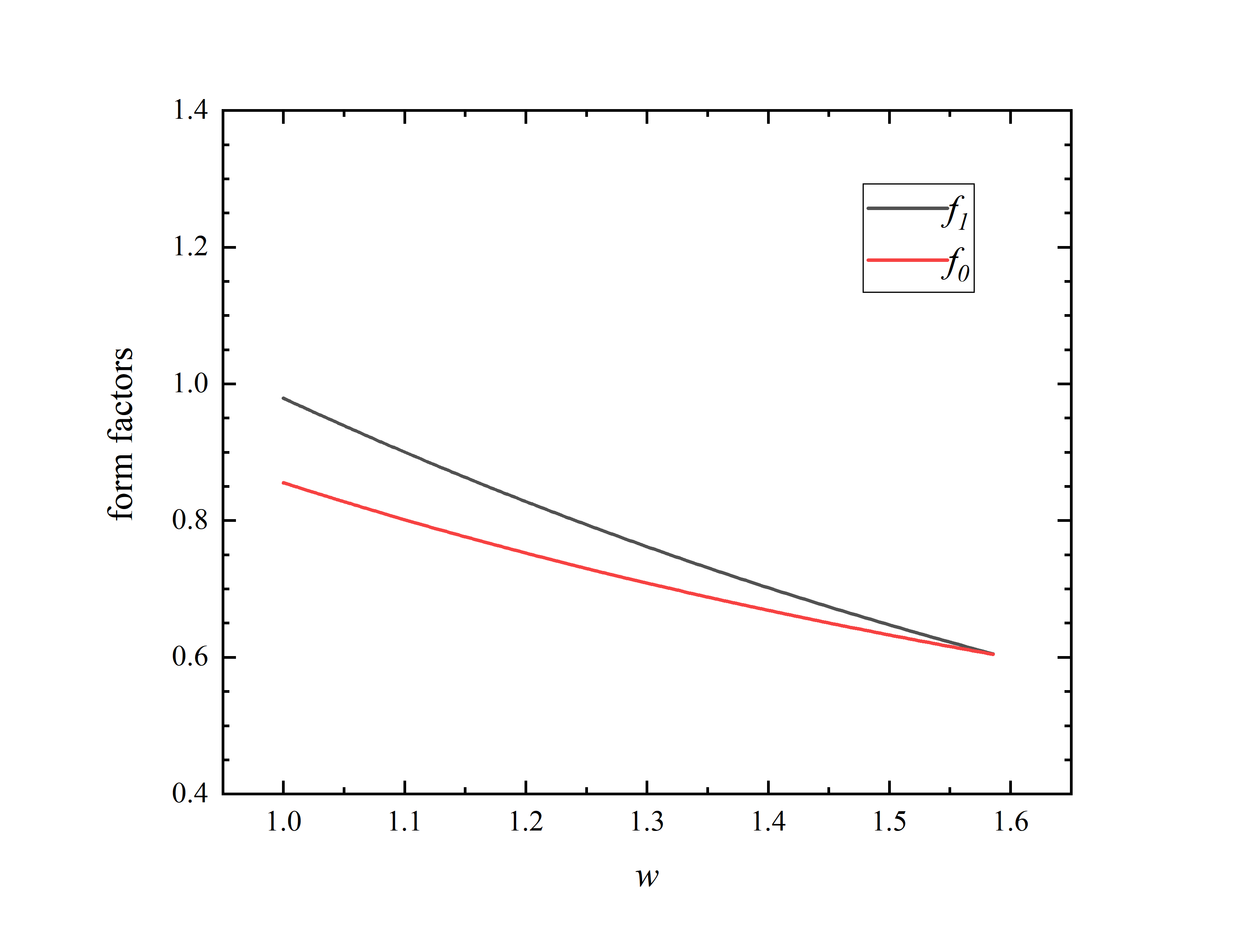}
   \subcaption{$f_1(w)$ and $f_0(w)$ without errors}
  \end{minipage}%
  \begin{minipage}[t]{0.5\textwidth}
   \centering
   \includegraphics[width=3.4in]{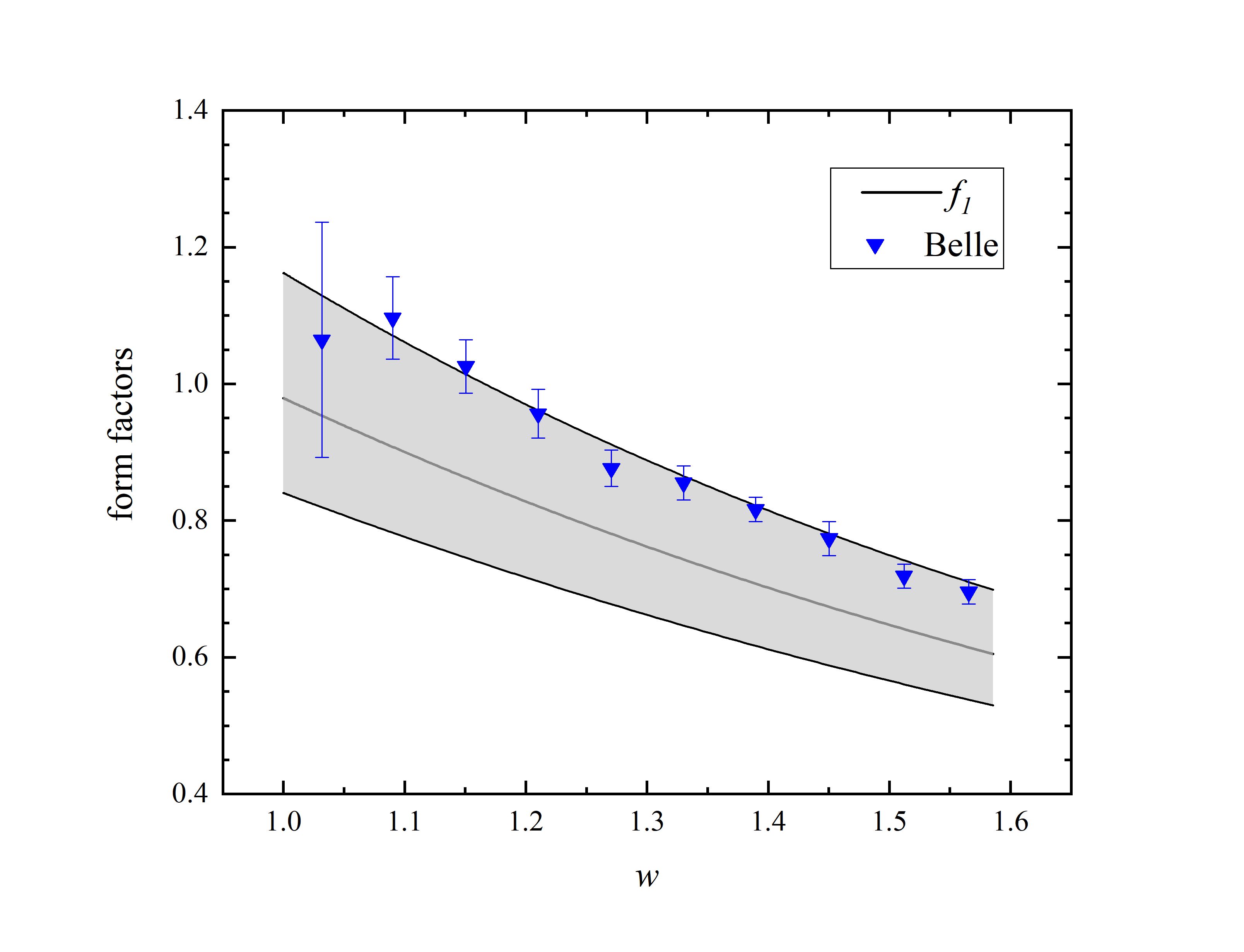}
   \subcaption{$f_1(w)$  with errors compared with Belle results \cite{Glattauer:2015teq}}
  \end{minipage}
  \caption{\label{fig:a2} The form factors with and without errors of the decay $\bar{B}^0\rightarrow D^{+}e\bar\nu_e$. }
\end{figure}

For the decay process $B^0 \rightarrow D^- \ell \overline{\nu}_{\ell}$, we can get \cite{Glattauer:2015teq}
\begin{equation}\label{eqgz}
\begin{split}
\mathcal{G}(z)&=\frac{2 \sqrt{r}}{(1+r)} f_{1}(w)
\\
&=\mathcal{G}(1)\left(1-8 \rho^{2} z+\left(51 \rho^{2}-10\right) z^{2}-\left(252 \rho^{2}-84\right) z^{3}\right),
\end{split}
\end{equation}
where
\begin{equation}\label{eqz}
z(w)=\frac{\sqrt{w+1}-\sqrt{2}}{\sqrt{w+1}+\sqrt{2}} .
\end{equation}
Here we have defined $r={m_D}/{m_B}$. If the lepton mass can be neglected, the differential decay rate will not depend on $f_{0}\left(w\right)$. Then we can express the form factors of channel $\bar{B}^0\rightarrow D^{+}ev_e$ as the functions of $w$, which is show in Fig.\ref{fig:a2}(a). In Fig.\ref{fig:a2}(b), we compare our result of $f_1$ with the experimental data~\cite{Glattauer:2015teq}, to show the uncertainty of the input parameters. We vary all the model parameters simultaneously around their center values within  $\pm10 \%$ and take the largest uncertainty as the errors. Within theoretical uncertainties, our results consist with Belle's data.

$\mathcal{G}(1)$ and $\rho^{2}$ are independent parameters which describe respectively the normalization and the shape of the measured decay distributions. Using Eq. \eqref{eqgz} and the numerical values of form factors as well as $|V_{c b}|=41.1 \times 10^{-3}$ from PDG \cite{Tanabashi:2018oca}, we obtain the values of $\mathcal{G}(1)$ and the slope $\rho^{2}$ which are shown in table \ref{tab:6}, where the average of experimental data and LQCD's results are also given as comparison. For both parameters, our results are smaller than the experimental data. However, considering the uncertainties, they are still consistent with each other. The normalization parameter $\mathcal{G}(1)$ and the slope $\rho^{2}$ for other semileptonic decays of $B_q$ meson to the pseudoscalar are shown in table \ref{tab:7}.

\begin{table}[htbp]
\centering
 \caption{\label{tab:6} The normalization and the slope of $B^0 \rightarrow D^- \ell \nu_\ell$.}.
 \begin{tabular}{cccc}
   \Xhline{1.5pt}
 parameters &  $\eta_{\mathrm{EW}} \mathcal{G}(1)|V_{c b}|[10^{-3}]$ & $\rho^{2}$  \\
  \midrule
ours & $35.6^{+4.9}_{-5.0}$  & $0.97\pm0.16$
\\
Averages of EX \cite{Amhis:2016xyh} &  $41.57 \pm 0.45_{\text { stat }} \pm 0.89_{\text { syst }}$  &  $1.128 \pm 0.024_{\text { stat }} \pm 0.023_{\text { syst }}$
\\
LQCD \cite{Na:2015kha} & 42.81(40)      &1.119(71)
\\
\bottomrule
 \end{tabular}
\end{table}

\begin{table}[htbp]
\centering
 \caption{\label{tab:7} The normalization and the slope of $B^-$, $B_s$ and $B_c$ decays to a pseudoscalar meson.}
 \begin{tabular}{cccc}
   \Xhline{1.5pt}
 Channel &  $\eta_{\mathrm{EW}} \mathcal{G}(1)|V_{c b}|[10^{-3}]$ & $\rho^{2}$  \\
  \midrule
$B^-\rightarrow D^{0}e\nu_e$ & $35.5^{+4.6}_{-4.8}$  & $0.96\pm0.13$
\\
$B_s\rightarrow D_s e\nu_e$ & $35.9^{+4.5}_{-4.7}$  & $1.13\pm0.21$
\\
$B_c\rightarrow \eta_c e\nu_e$ & $37.4^{+4.8}_{-4.2}$  & $2.64\pm0.23$
\\
\bottomrule
 \end{tabular}
\end{table}

For the decay process $\bar{B}^0 \rightarrow D^{*+} \ell \overline{\nu}_{\ell}$, where the final meson is a vector, we can get a similar formula as that of the pseudoscalar case \cite{Harrison:2017fmw}
\begin{equation}
\frac{d \Gamma\left(\bar{B}^0 \rightarrow D^{*+} \ell \overline{\nu}_{\ell}\right)}{d w}=\frac{G_{\mathrm{F}}^{2} m_{D^{*}}^{3}}{48 \pi^{3}}\left(m_{B}-m_{D^{*}}\right)^{2} \eta_{\mathrm{EW}}^{2} \chi(w) \mathcal{F}^{2}(w)\left|V_{c b}\right|^{2},
\end{equation}
where
\begin{equation}
\begin{array}{l}{\chi(w) \mathcal{F}^{2}(w)=} \\ {\qquad \begin{array}{l}{h_{A_{1}}^{2}(w) \sqrt{w^{2}-1}(w+1)^{2}\left\{2\left[\frac{1-2 w r+r^{2}}{(1-r)^{2}}\right]\left[1+R_{1}^{2}(w) \frac{w^{2}-1}{w+1}\right]+\right.} \\ {\left[1+\left(1-R_{2}(w)\right) \frac{w-1}{1-r}\right]^{2} \},}\end{array}}\end{array}
\end{equation}
and
\begin{equation}
\begin{split}\label{eqhw}
h_{A_{1}}(w)&=\frac{2 \sqrt{m_{B} m_{D^{*}}}}{m_{B}+m_{D^{*}}} \frac{A_{1}\left(Q^{2}\right)}{1-\frac{Q^{2}}{\left(m_{B}+m_{D^{*}}\right)^{2}}}.
\end{split}
\end{equation}
Here we use the parametrization
of form factors  introduced by Caprini, Lellouch and Neubert (CLN) \cite{Caprini:1997mu}
\begin{equation}\label{eqhr}
\begin{aligned} h_{A_{1}}(w) &=h_{A_{1}}(1)\left[1-8 \rho^{2} z+\left(53 \rho^{2}-15\right) z^{2}-\left(231 \rho^{2}-91\right) z^{3}\right], \\ R_{1}(w) & =\frac{h_{V}(w)}{h_{A_{1}}(w)} \approx 1.27-0.12(w-1)+0.05(w-1)^{2}, \\ R_{2}(w)&=\frac{h_{A_{3}}(w)+r h_{A_{2}}(w)}{h_{A_{1}}(w)} \approx 0.80+0.11(w-1)-0.06(w-1)^{2}. \end{aligned}
\end{equation}
All these functions, $h_{A_{1}}(w)$, $R_{1}(w)$, $R_{2}(w)$ etc., are alterations of the former form factors, which can be obtained by using Eq. \eqref{e8}. For example, at zero recoil, where $w=1$ , $z=0$ and $Q^{2}=Q_{\max }^{2}=(m_B+m_D)^2$, we have
\begin{equation}
\mathcal{F}(1)=h_{A_{1}}(1)=\frac{m_{B}+m_{D^{*}}}{2 \sqrt{m_{B} m_{D^{*}}}} A_{1}\left(Q_{\max }^{2}\right).
\end{equation}
We will not show details of other functions, but present the the results of  $\eta_{\mathrm{EW}} \mathcal{F}(1)\left|V_{c b}\right|$, $\rho^2$, $R_{1}(1)$ and $R_{2}(1)$ in table \ref{tab:8}, where the LQCD results and the averages of experimental data are also listed for comparison.
\begin{table}[htbp]
\centering
 \caption{\label{tab:8} The normalization and the slope of $B^0 \rightarrow D^{*-} \ell \nu_\ell$.}
 \begin{tabular}{ccccccc}
   \Xhline{1.5pt}
 parameters &  $\eta_{\mathrm{EW}} \mathcal{F}(1)|V_{c b}|[10^{-3}]$ & $\rho^{2}$\\
  \midrule
ours & $40.06^{+4.98}_{-4.13}$  & $1.04\pm0.19$
\\
Averages of EX \cite{Amhis:2016xyh} &  $35.61 \pm 0.11_{\text { stat }} \pm 0.41_{\text { syst }}$  &  $1.205 \pm 0.015_{\mathrm{stat}} \pm 0.021_{\mathrm{syst}}$
\\
LQCD \cite{Harrison:2017fmw} & $36.71\pm0.41\pm0.84$      & 1.29(17)
\\
\Xhline{1.5pt}
parameters & $R_1(1)$ & $R_2(1)$  \\
\midrule
ours & $1.55^{+0.12}_{-0.11}$ & $0.98^{+0.12}_{-0.11}$
\\
LQCD \cite{Harrison:2017fmw} & $1.404\pm0.032$ & $0.854\pm0.020$
\\
\bottomrule
 \end{tabular}
\end{table}
Similarly, in table \ref{tab:9}, the normalization $\mathcal{G}(1)$ and the slope $\rho^{2}$ of the other $0^- \rightarrow 1^-$ channels are shown.
\begin{table}[htbp]
\centering
 \caption{\label{tab:9} The normalization and the slope of $B^-$, $B_s$ and $B_c$ decay to a vector meson.}
 \begin{tabular}{cccccc}
   \Xhline{1.5pt}
 Channel &  $\eta_{\mathrm{EW}} \mathcal{F}(1)|V_{c b}|[10^{-3}]$ & $\rho^{2}$ & $R_1(1)$ & $R_2(1)$ \\
  \midrule
$B^-\rightarrow D^{*0}ev_e$ & $40.1^{+4.9}_{-4.2}$  & $1.04\pm0.18$ & $1.55^{+0.11}_{-0.12}$ & $0.98^{+0.12}_{-0.11}$
\\
$B_s\rightarrow D_s^* ev_e$ & $39.8^{+4.6}_{-4.4}$  & $1.18\pm0.15$ & $1.34^{+0.10}_{-0.11}$ & $1.03^{+0.15}_{-0.14}$
\\
$B_c\rightarrow J/\psi ev_e$ & $38.8^{+4.8}_{-4.1}$  & $2.67\pm0.16$
\\
\bottomrule
 \end{tabular}
\end{table}

\begin{figure}[h]
  \begin{minipage}[t]{0.5\textwidth}
   \centering
   \includegraphics[width=3.4in]{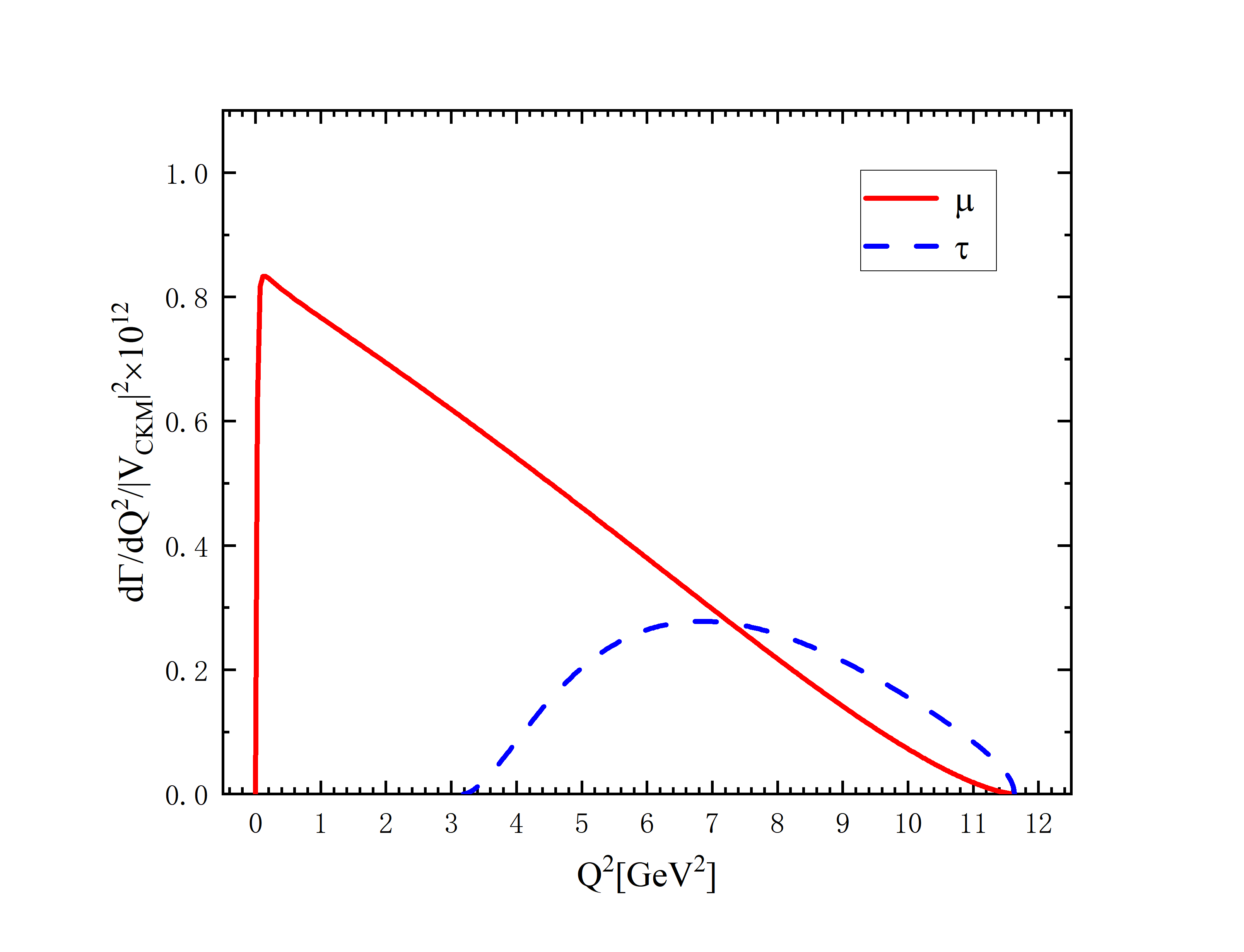}
   \subcaption{The spectrum for $B \rightarrow D \mu( \tau ) \nu $}
   \label{fig:side:e}
  \end{minipage}%
  \begin{minipage}[t]{0.5\textwidth}
   \centering
   \includegraphics[width=3.4in]{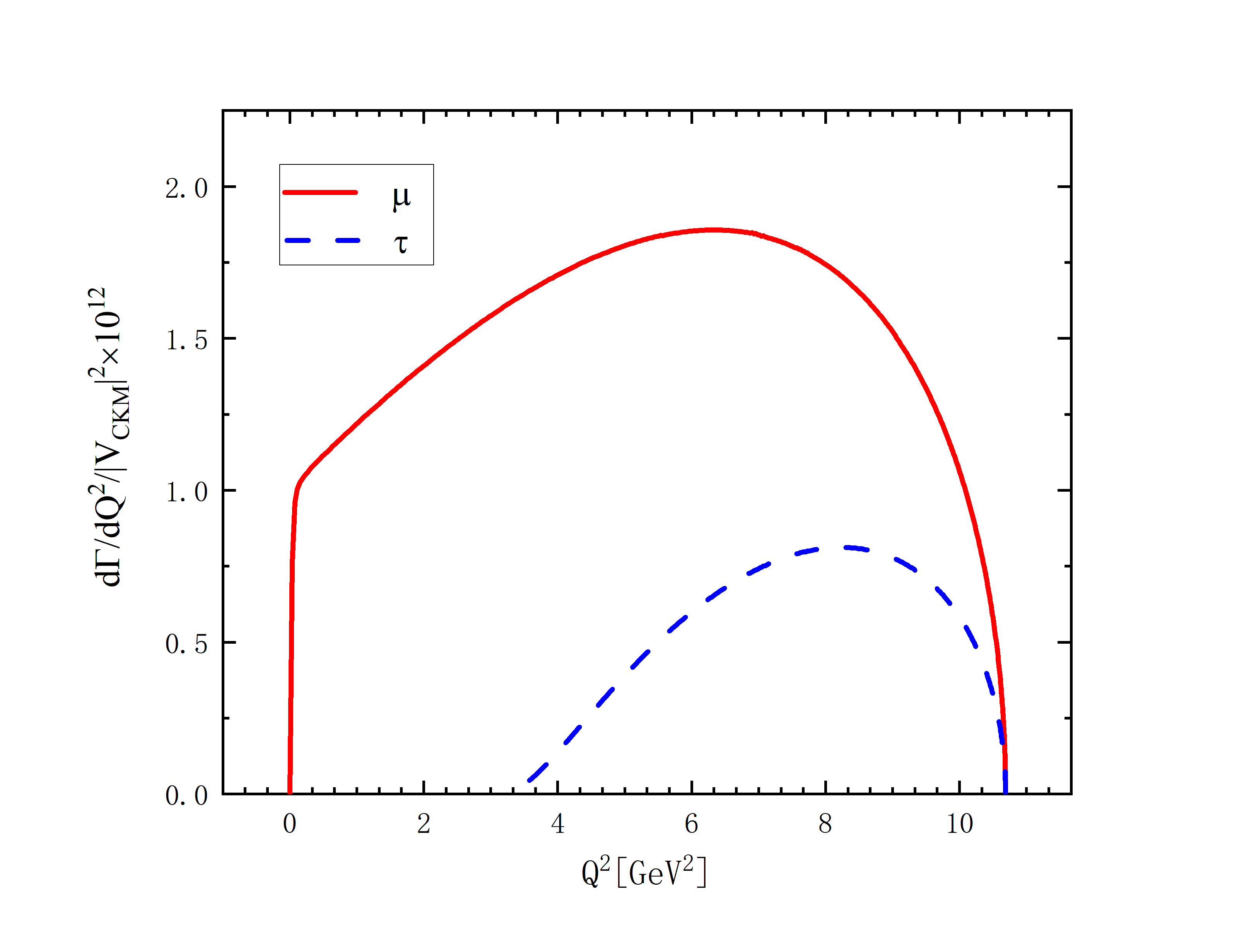}
   \subcaption{The spectrum for $B \rightarrow D^{*} \mu( \tau ) \nu $}
   \label{fig:side:f}
  \end{minipage}
  \caption{\label{fig:a6} The differential branching fractions for B meson decays.}
\end{figure}

\begin{figure}[h]
  \begin{minipage}[t]{0.5\textwidth}
   \centering
   \includegraphics[width=3.4in]{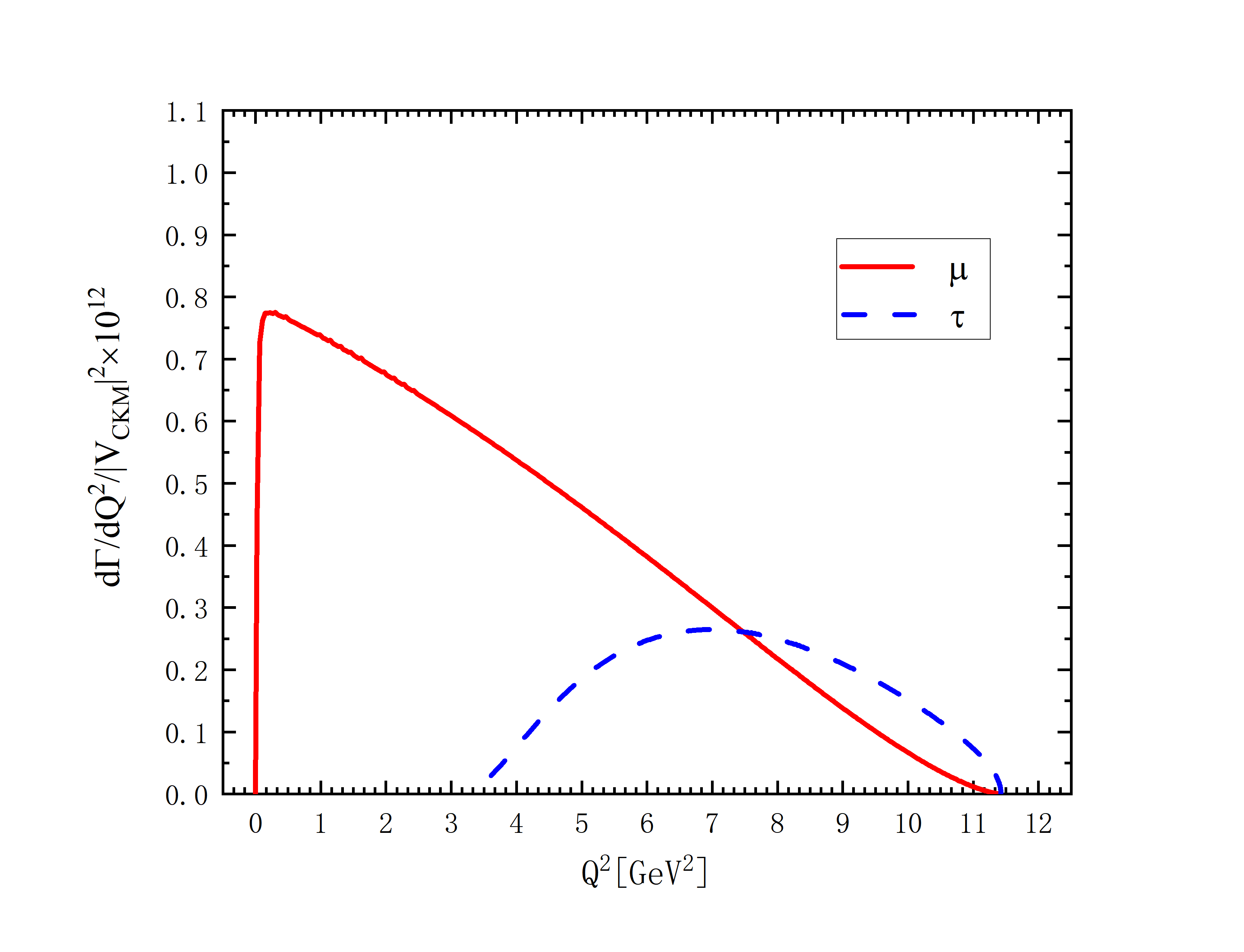}
   \subcaption{The spectrum for $B_s \rightarrow D_s \mu( \tau ) \nu $}
   \label{fig:side:e}
  \end{minipage}%
  \begin{minipage}[t]{0.5\textwidth}
   \centering
   \includegraphics[width=3.4in]{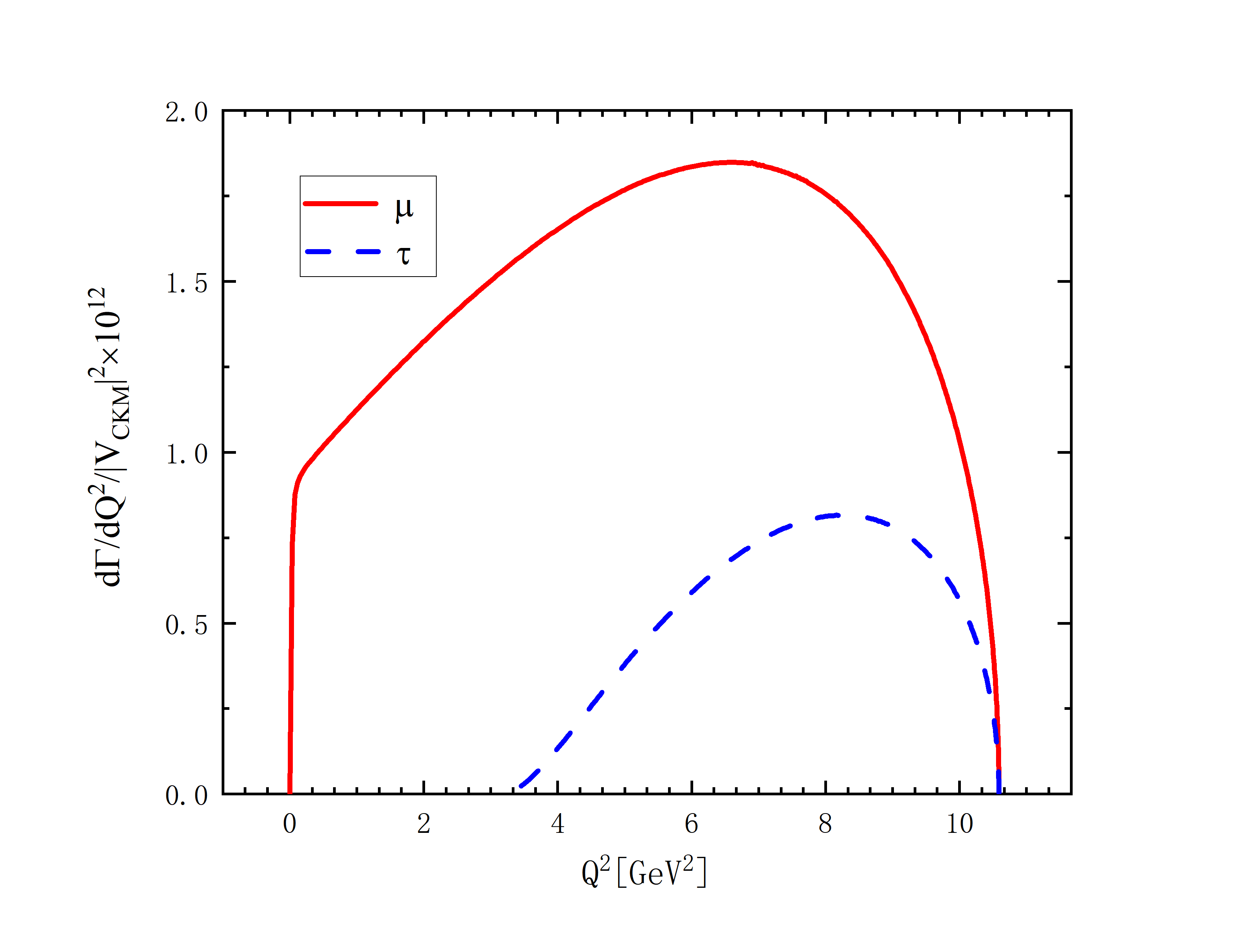}
   \subcaption{The spectrum for $B_s \rightarrow D_s^{*} \mu( \tau ) \nu $}
   \label{fig:side:f}
  \end{minipage}
  \caption{\label{fig:a7}The differential branching fractions for $B_s$ meson decays.}
\end{figure}

\begin{figure}[h]
  \begin{minipage}[t]{0.5\textwidth}
   \centering
   \includegraphics[width=3.4in]{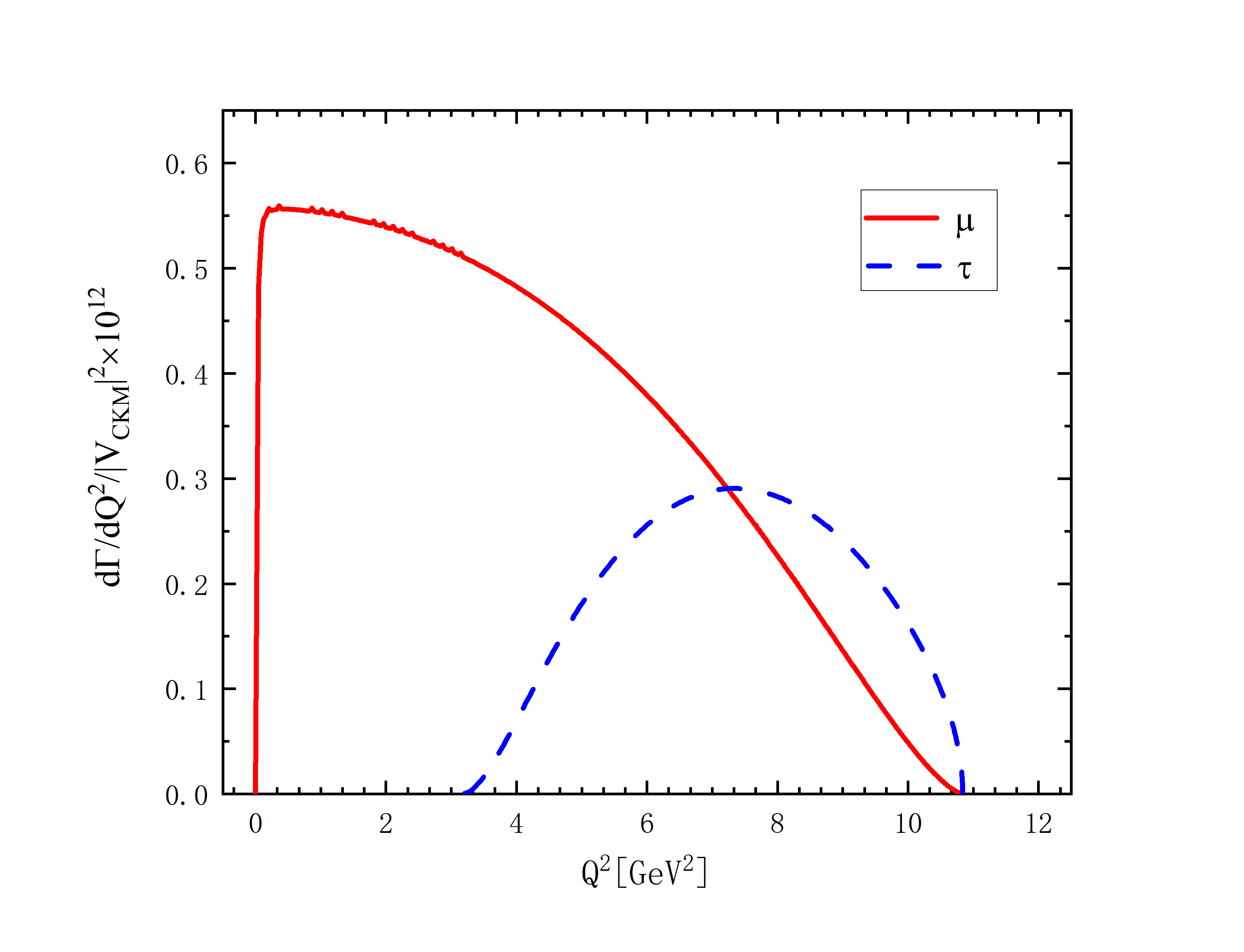}
   \subcaption{The spectrum for $B_c \rightarrow \eta_c \mu( \tau ) \nu $}
   \label{fig:side:e}
  \end{minipage}%
  \begin{minipage}[t]{0.5\textwidth}
   \centering
   \includegraphics[width=3.4in]{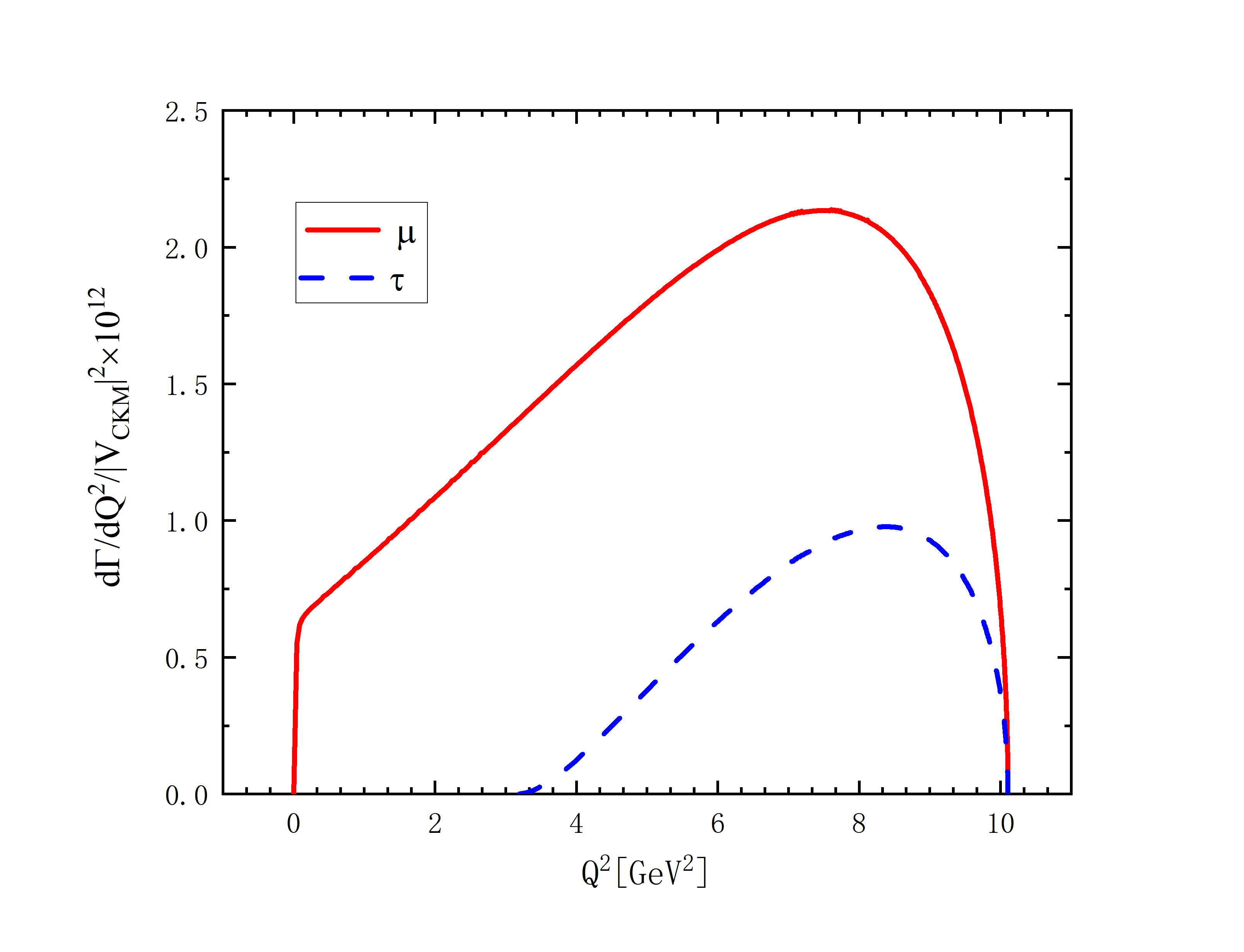}
   \subcaption{The spectrum for $B_c \rightarrow J/\psi \mu( \tau ) \nu $}
   \label{fig:side:f}
  \end{minipage}
  \caption{\label{fig:a8} The differential branching fractions for $B_c$ meson decays. }
\end{figure}

The differential branching fraction is another observable. With the numerical values of form factors calculated by the improved BS method, we straightforwardly obtain differential branching fractions. In Fig.\ref{fig:a6}(a), the spectra of $B \to D \mu\nu$ and $B \to D \tau \nu$ are shown. In Fig.\ref{fig:a6}(b), the spectra for $B \to D^* \mu\nu$ and $B \to D^* \tau \nu$ are given. Our results of differential branching fractions for the cases of $B \to D$ agree  very well with those of Ref.~\cite{Harrison:2017fmw} by the lattice QCD method. In
Fig.\ref{fig:a7} and Fig.\ref{fig:a8}, we present respectively the differential branching fractions for $B_s$ and $B_c$ decays.

\begin{figure}[h]
  \begin{minipage}[t]{0.5\textwidth}
   \centering
   \includegraphics[width=3.4in]{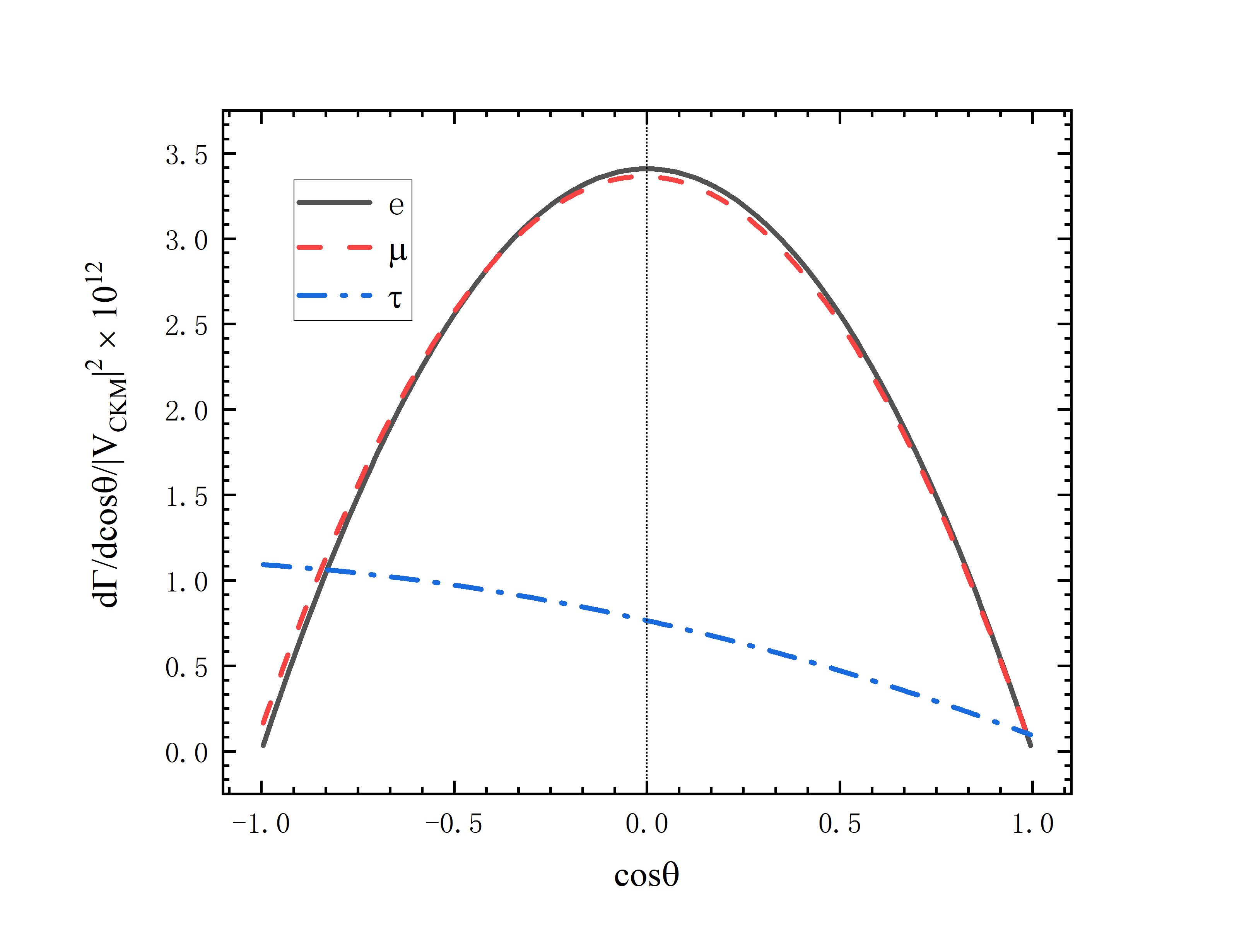}
   \subcaption{The spectrum for $B \rightarrow D \ell \nu_\ell $}
   \label{fig:side:e}
  \end{minipage}%
  \begin{minipage}[t]{0.5\textwidth}
   \centering
   \includegraphics[width=3.4in]{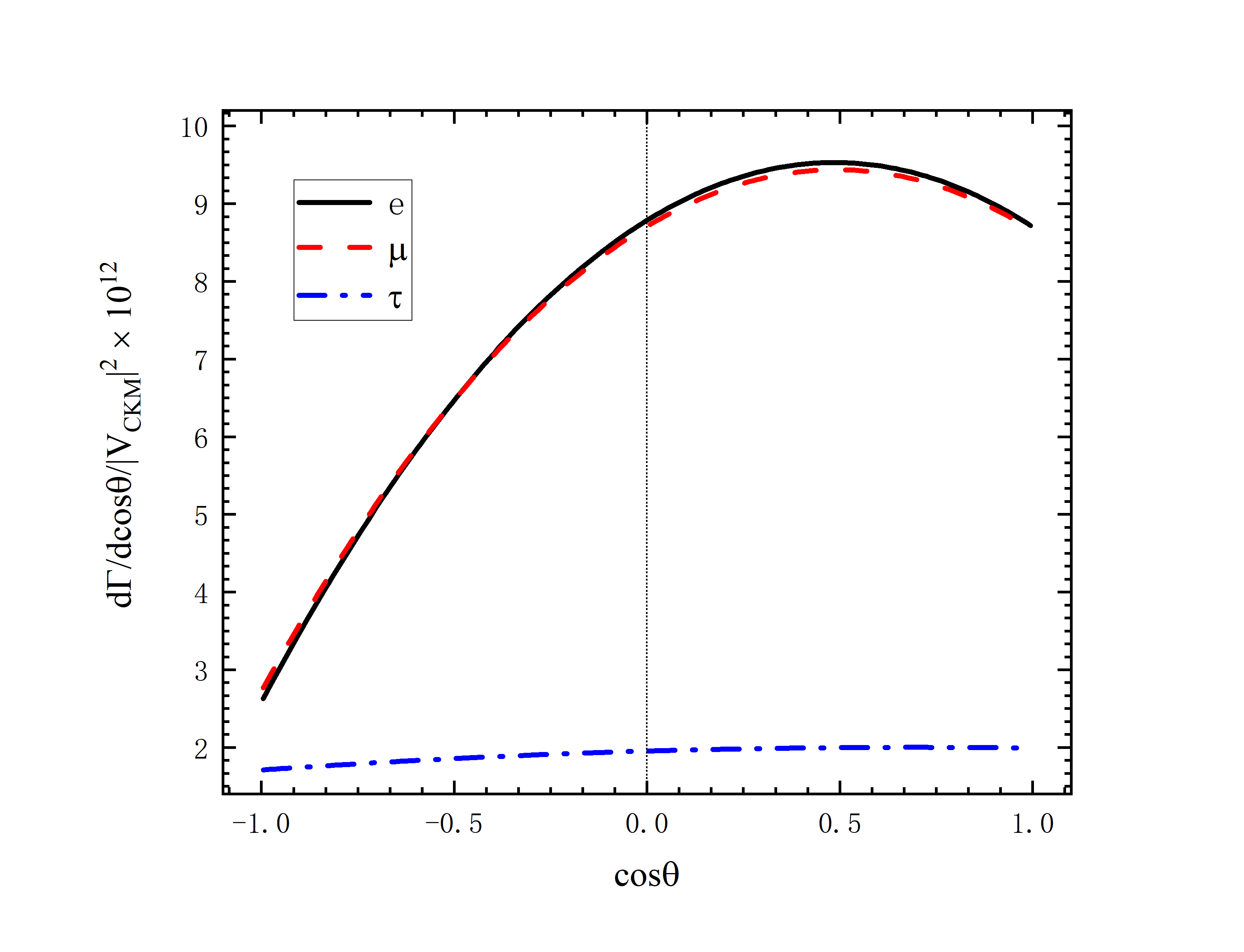}
   \subcaption{The spectrum for $B \rightarrow D^{*} \ell \nu_\ell $}
   \label{fig:side:f}
  \end{minipage}
  \caption{\label{fig:xspecB} The angular distributions for B meson decays.}
\end{figure}

\begin{figure}[h]
  \begin{minipage}[t]{0.5\textwidth}
   \centering
   \includegraphics[width=3.4in]{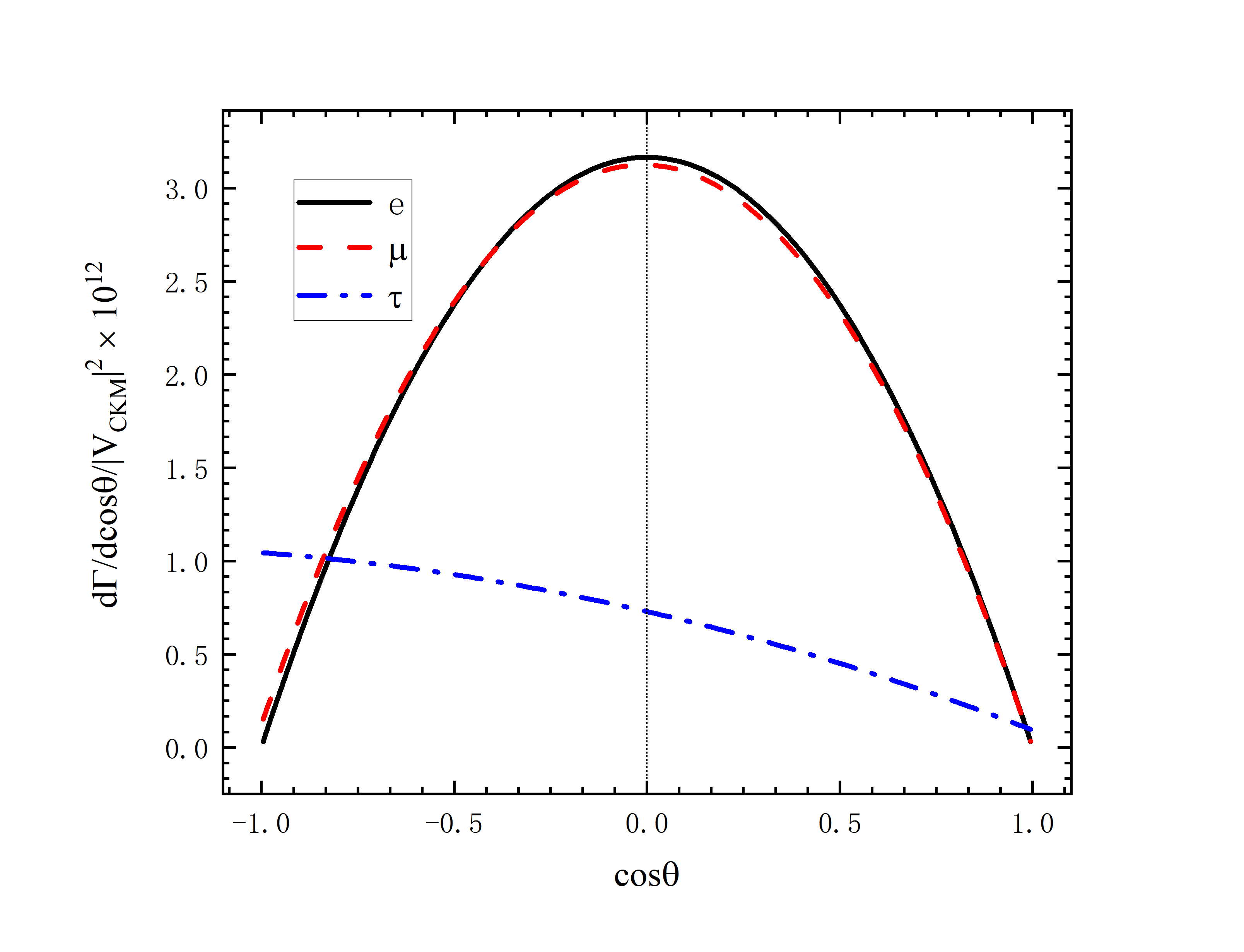}
   \subcaption{The spectrum for $B_s \rightarrow D_s \ell \nu_\ell $}
   \label{fig:side:e}
  \end{minipage}%
  \begin{minipage}[t]{0.5\textwidth}
   \centering
   \includegraphics[width=3.4in]{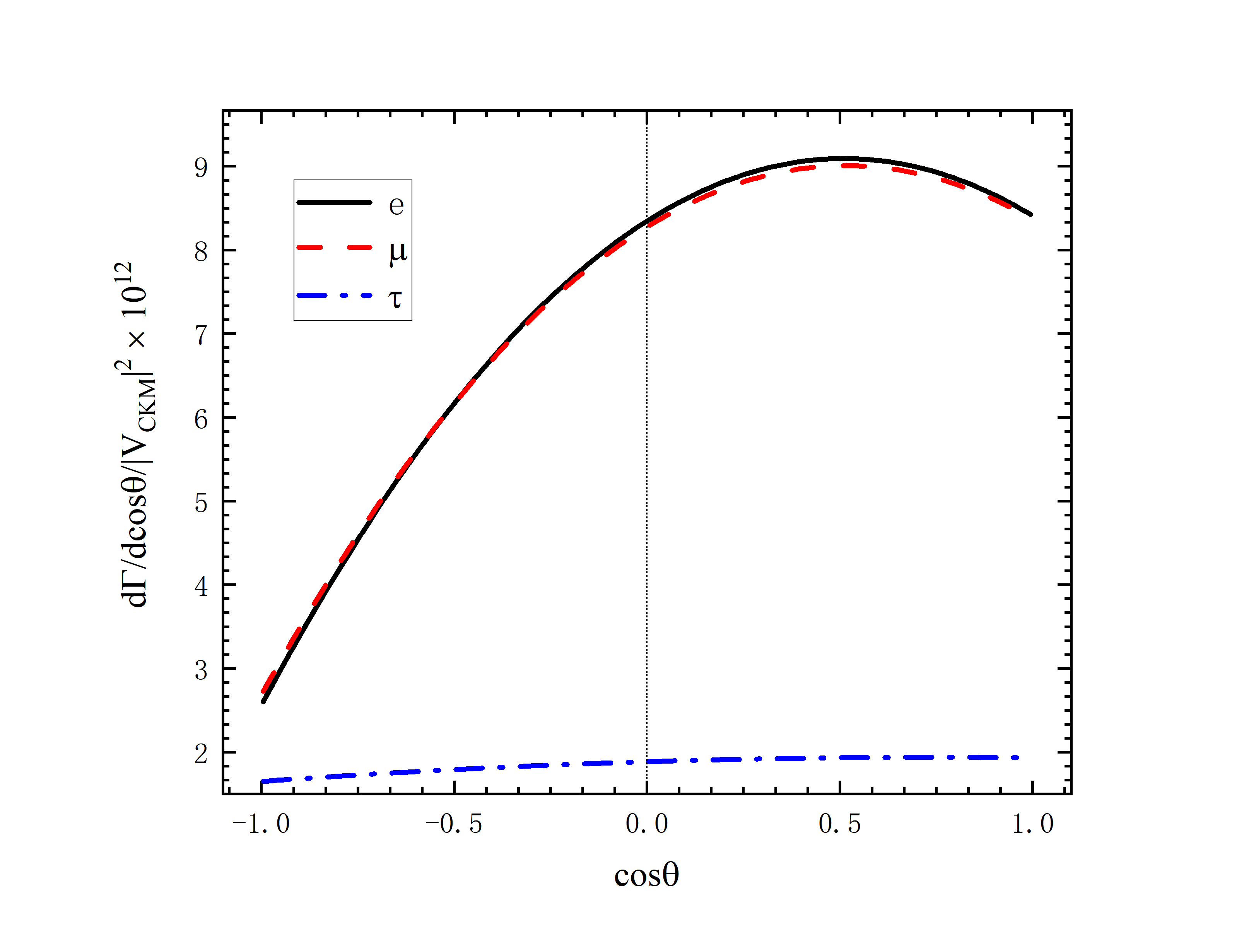}
   \subcaption{The spectrum for $B_s \rightarrow D_s^{*} \ell \nu_\ell $}
   \label{fig:side:f}
  \end{minipage}
  \caption{\label{fig:xspecBs}The angular distributions for $B_s$ meson decays.}
\end{figure}

\begin{figure}[h]
  \begin{minipage}[t]{0.5\textwidth}
   \centering
   \includegraphics[width=3.4in]{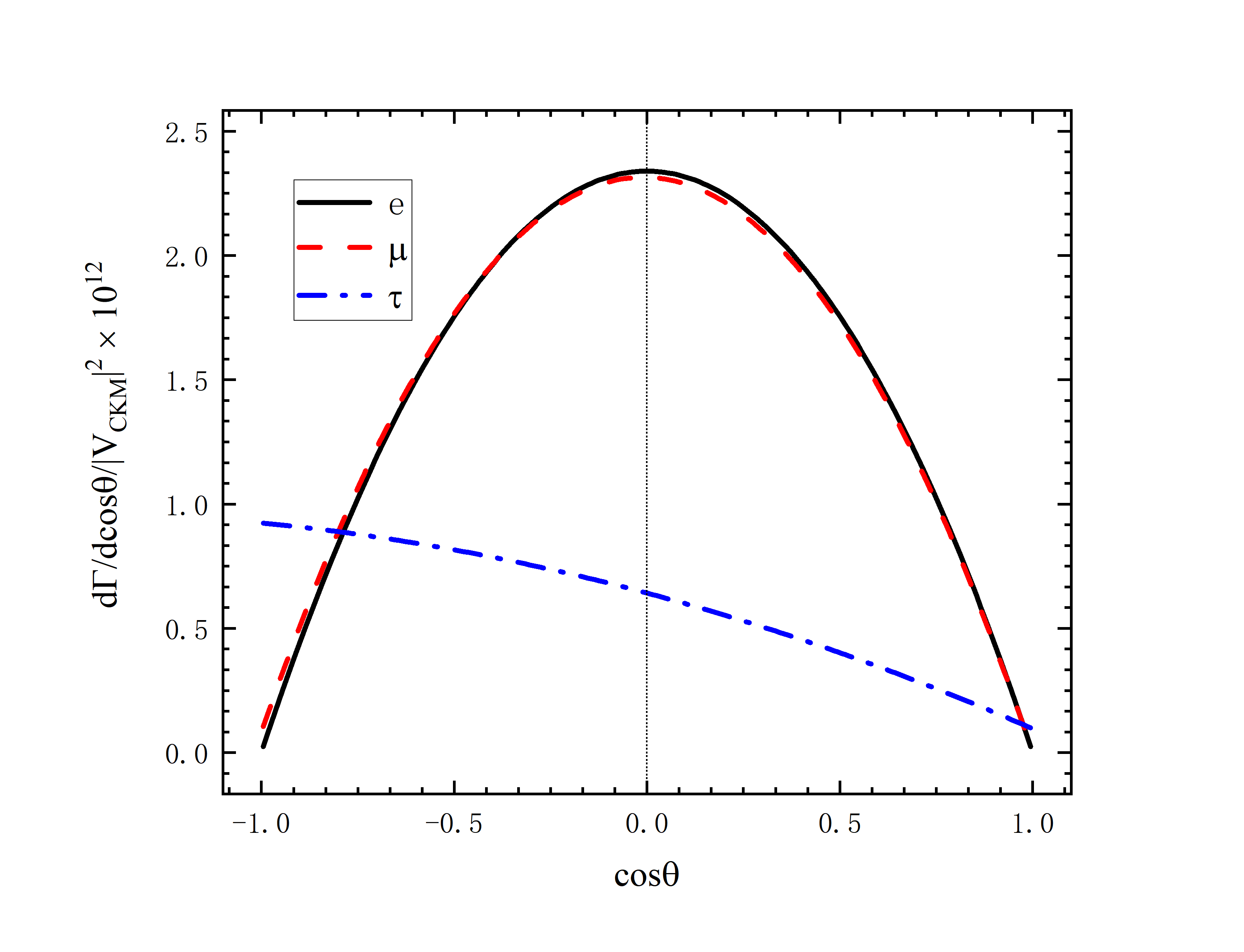}
   \subcaption{The spectrum for $B_c \rightarrow \eta_c \ell \nu_\ell $}
   \label{fig:side:e}
  \end{minipage}%
  \begin{minipage}[t]{0.5\textwidth}
   \centering
   \includegraphics[width=3.4in]{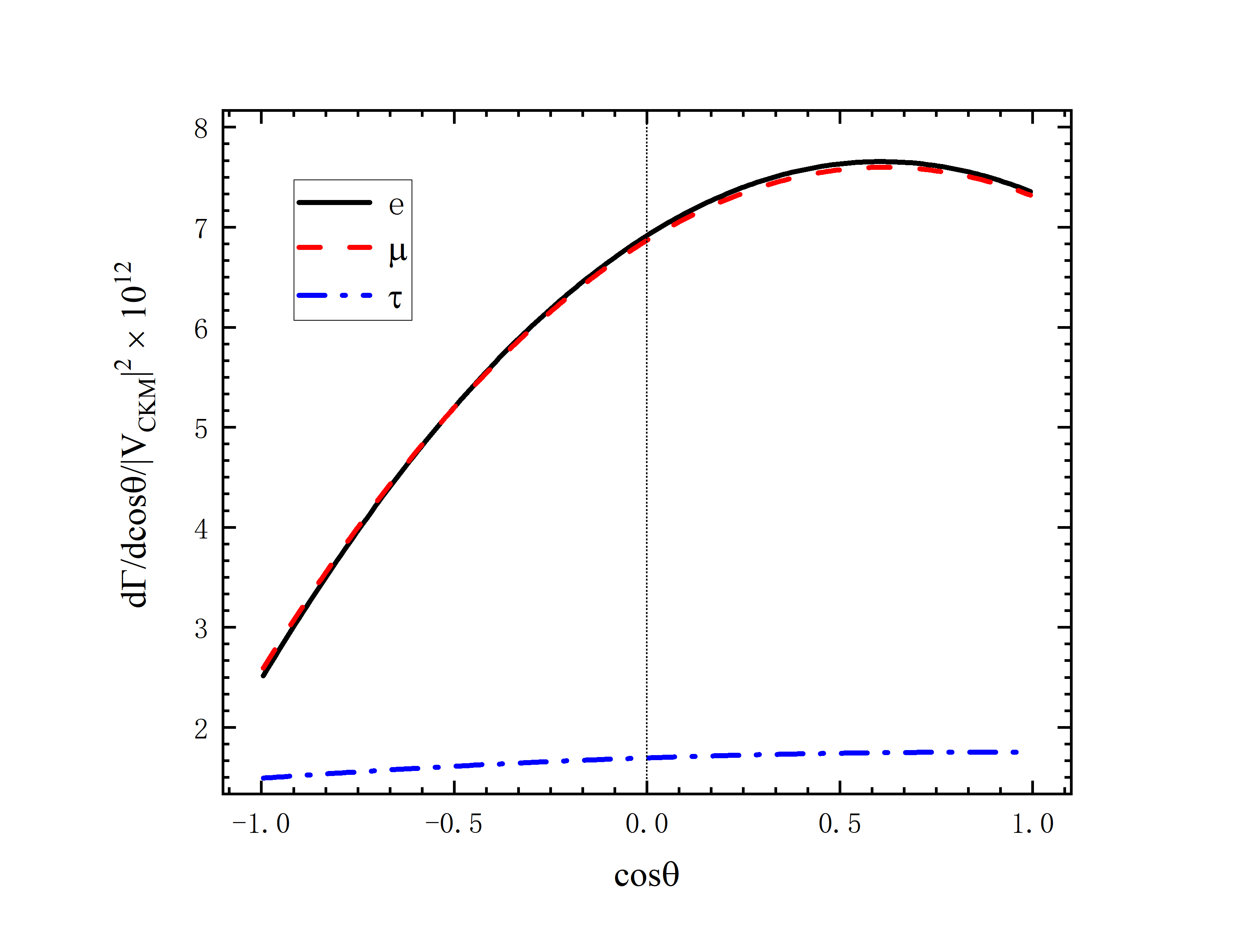}
   \subcaption{The spectrum for $B_c \rightarrow J/\psi \ell \nu_\ell $}
   \label{fig:side:f}
  \end{minipage}
  \caption{\label{fig:xspecBc} The angular distributions for $B_c$ meson decays.}
\end{figure}

\begin{figure}[h]
  \begin{minipage}[t]{0.5\textwidth}
   \centering
   \includegraphics[width=3.4in]{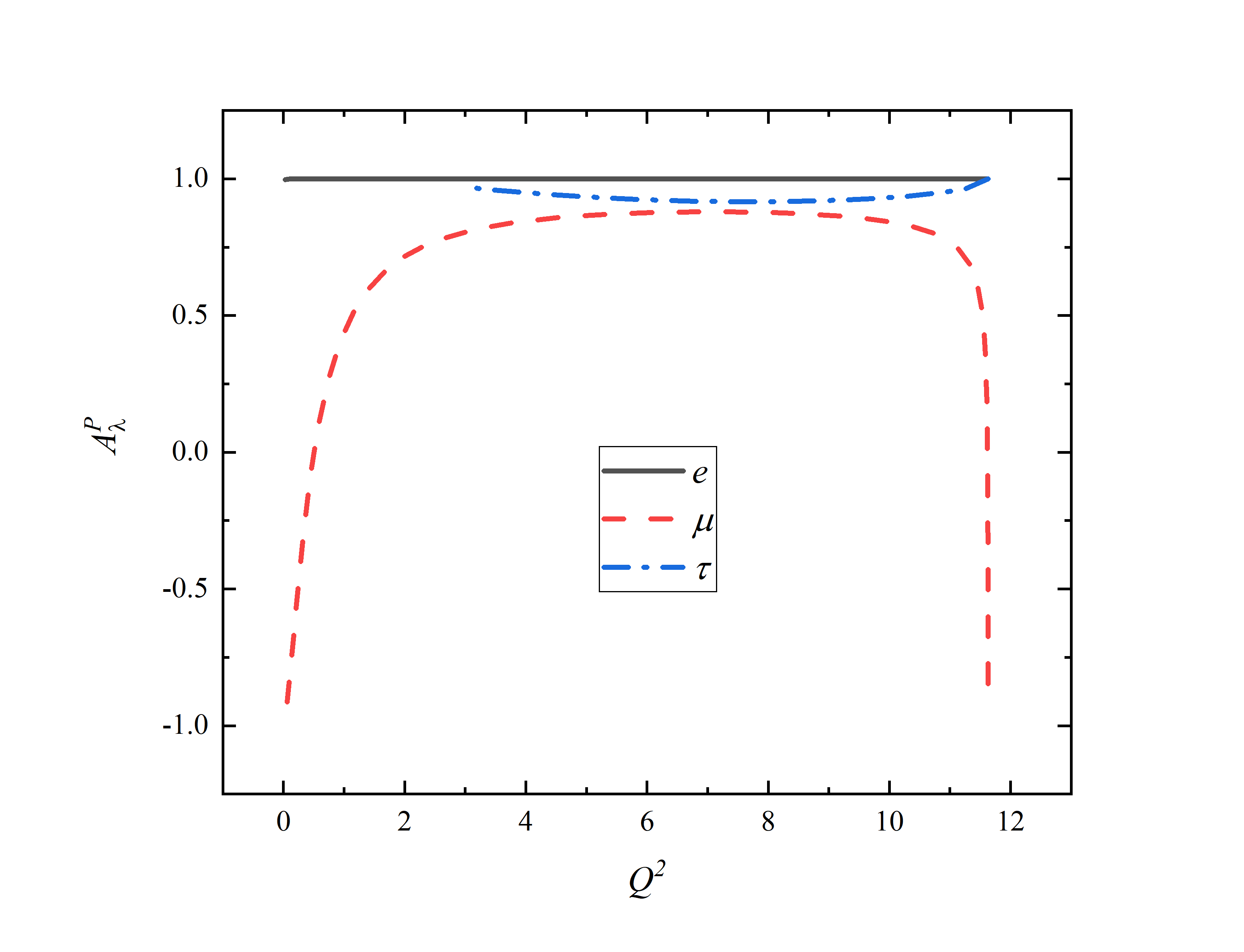}
  \end{minipage}%
  \begin{minipage}[t]{0.5\textwidth}
   \centering
   \includegraphics[width=3.4in]{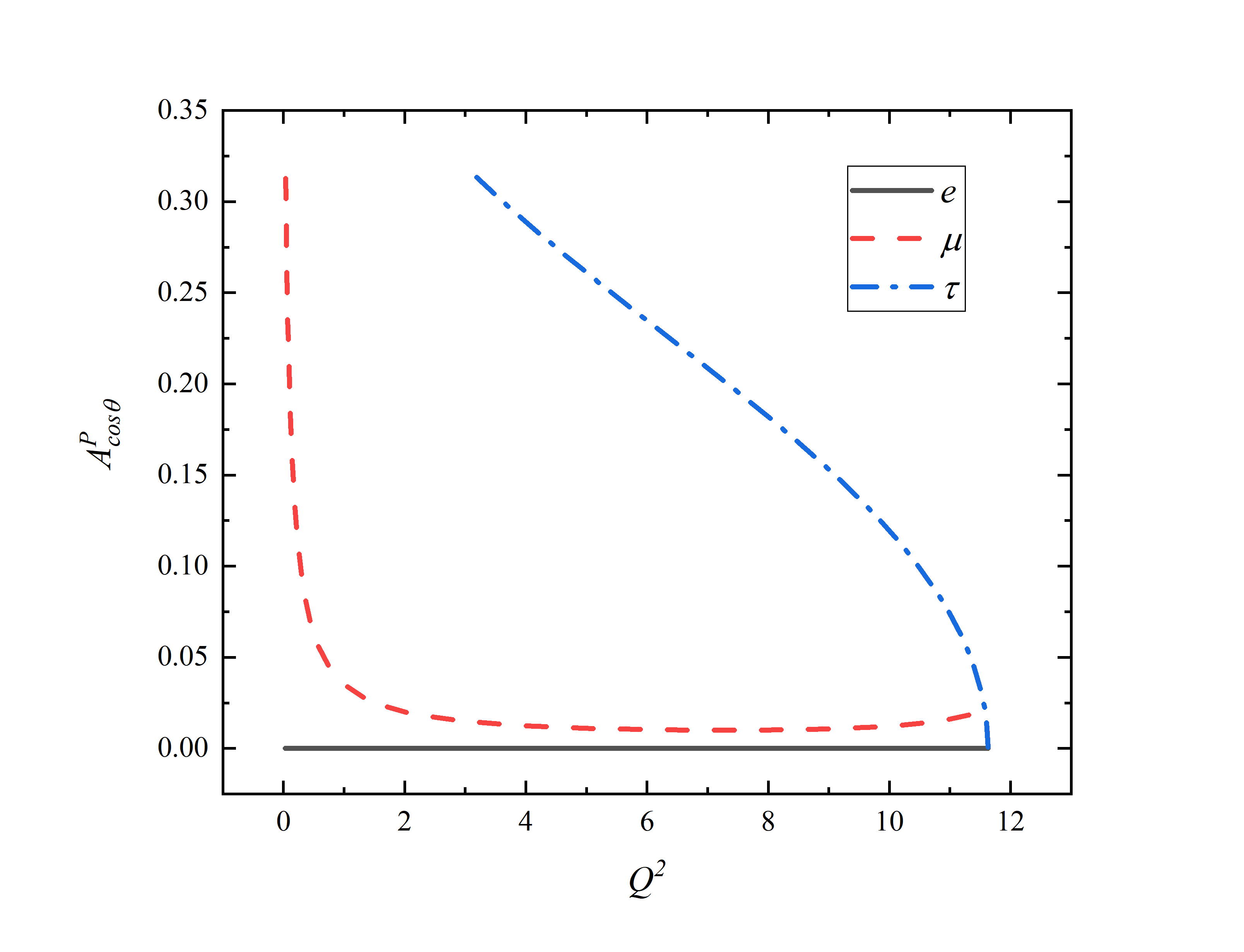}
  \end{minipage}
  \caption{\label{fig:B0alac} $A_{\lambda}^{P}$ (left) and $A_{cos\theta}^{P}$(right) of the decay $\bar{B}^0\rightarrow D^{+}(1S)\ell \nu_\ell$.}
\end{figure}

\begin{figure}[h]
  \begin{minipage}[t]{0.5\textwidth}
   \centering
   \includegraphics[width=3.4in]{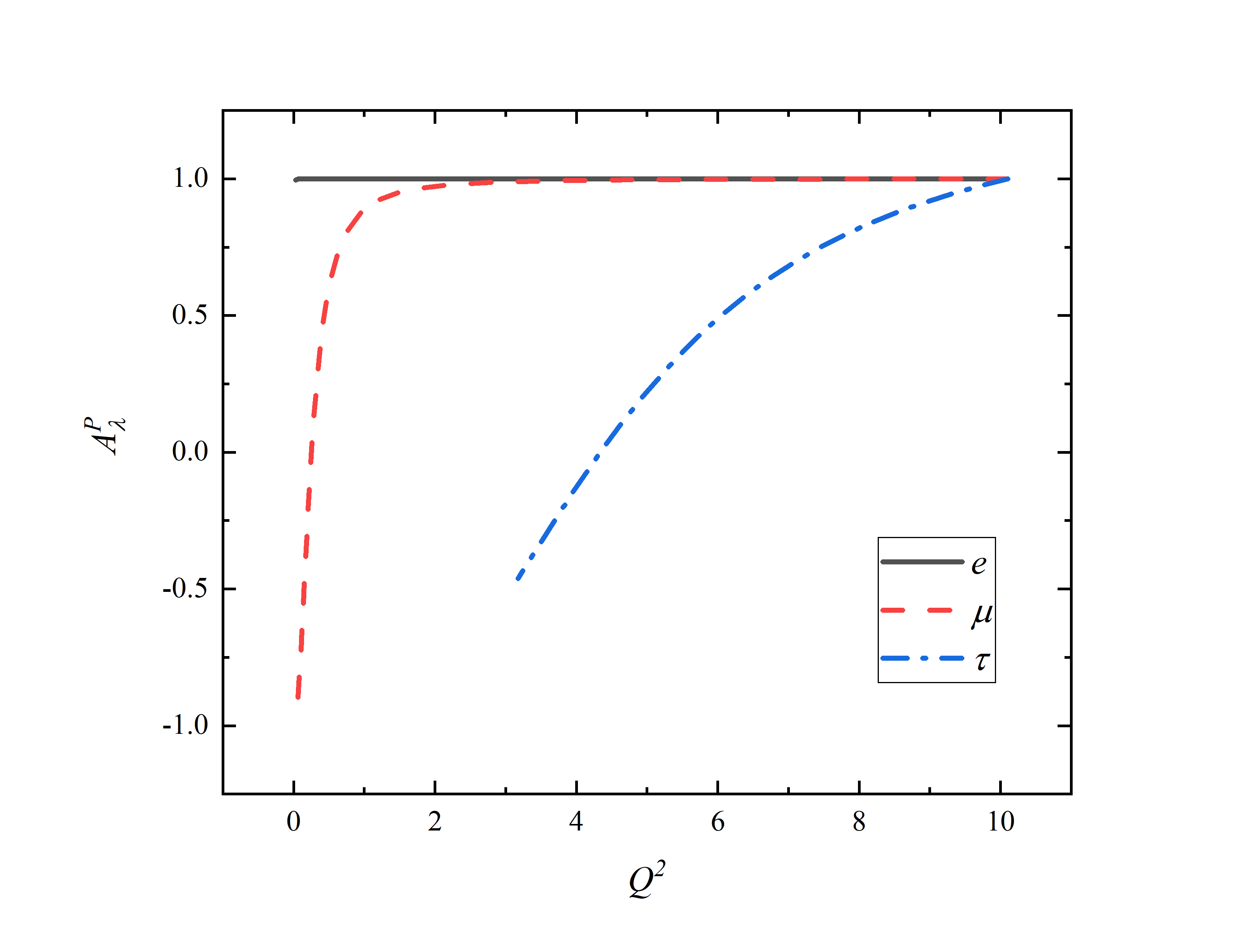}
  \end{minipage}%
  \begin{minipage}[t]{0.5\textwidth}
   \centering
   \includegraphics[width=3.4in]{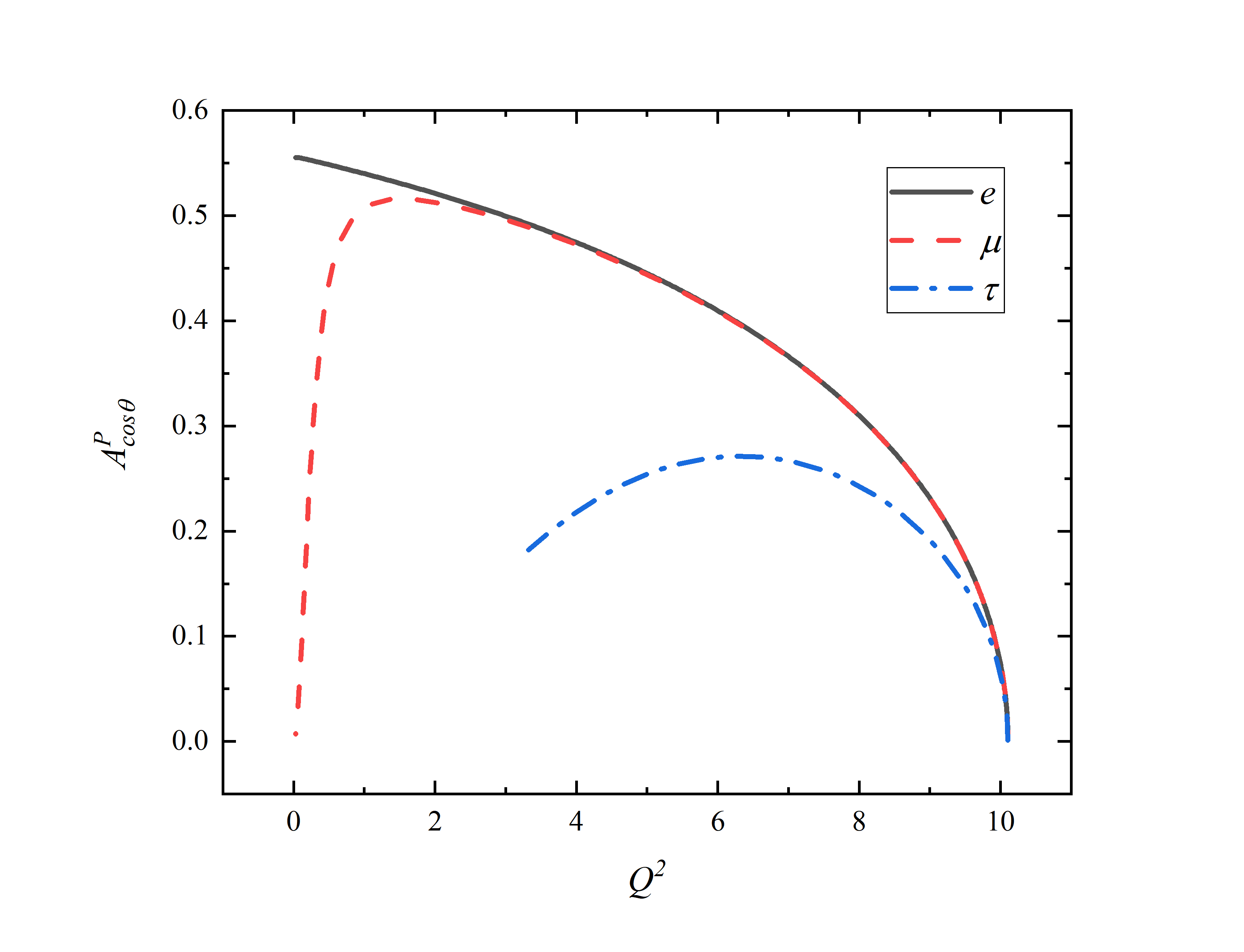}
  \end{minipage}
  \caption{\label{fig:Bcjalac} $A_{\lambda}^{P}$(left) and $A_{cos\theta}^{P}$(right) of the decay $B^-_c \rightarrow J/\psi \ell \nu_\ell$.}
\end{figure}

For completeness, we also calculate the angular distribution ${\rm d}\Gamma/{\rm dcos}\theta$, where $\theta$ is the angle between $\vec P_f$ (the momentum of the final meson) and $\vec P_l^\ast$ (the momentum of the charged lepton in the center-of-momentum frame of $l^-\bar\nu_l$). The results are presented in Fig. \ref{fig:xspecB}, Fig.\ref{fig:xspecBs}, and Fig.\ref{fig:xspecBc}.
Moreover, we can also study the the lepton spin asymmetry and the forward-backward asymmetry, which are defined respectively as
\begin{equation}
\begin{aligned}
A_{\lambda}^{P}\left(Q^{2}\right)=& \frac{d \Gamma\left[\lambda_{\ell}=-1 / 2\right] / d Q^{2}-d \Gamma\left[\lambda_{\ell}=1 / 2\right] / d Q^{2}}{d \Gamma\left[\lambda_{\ell}=-1 / 2\right] / d Q^{2}+d \Gamma\left[\lambda_{\ell}=1 / 2\right] / d Q^{2}},
\end{aligned}
\end{equation}
and 
\begin{equation}
\begin{aligned}
A_{cos\theta}^{P}\left(Q^{2}\right) = & \frac{\int_{-1}^{0} d \cos \theta\left(d^{2} \Gamma / d Q^{2} d \cos \theta\right)-\int_{0}^{1} d \cos \theta\left(d^{2} \Gamma / d Q^{2} d \cos \theta\right)}{d^{2} \Gamma / d Q^{2}}.
\end{aligned}
\end{equation}
The results are shown in Fig.\ref{fig:B0alac} $\sim$ Fig.\ref{fig:Bcjalac}.

\begin{figure}[h]
  \begin{minipage}[t]{0.5\textwidth}
   \centering
   \includegraphics[width=3.4in]{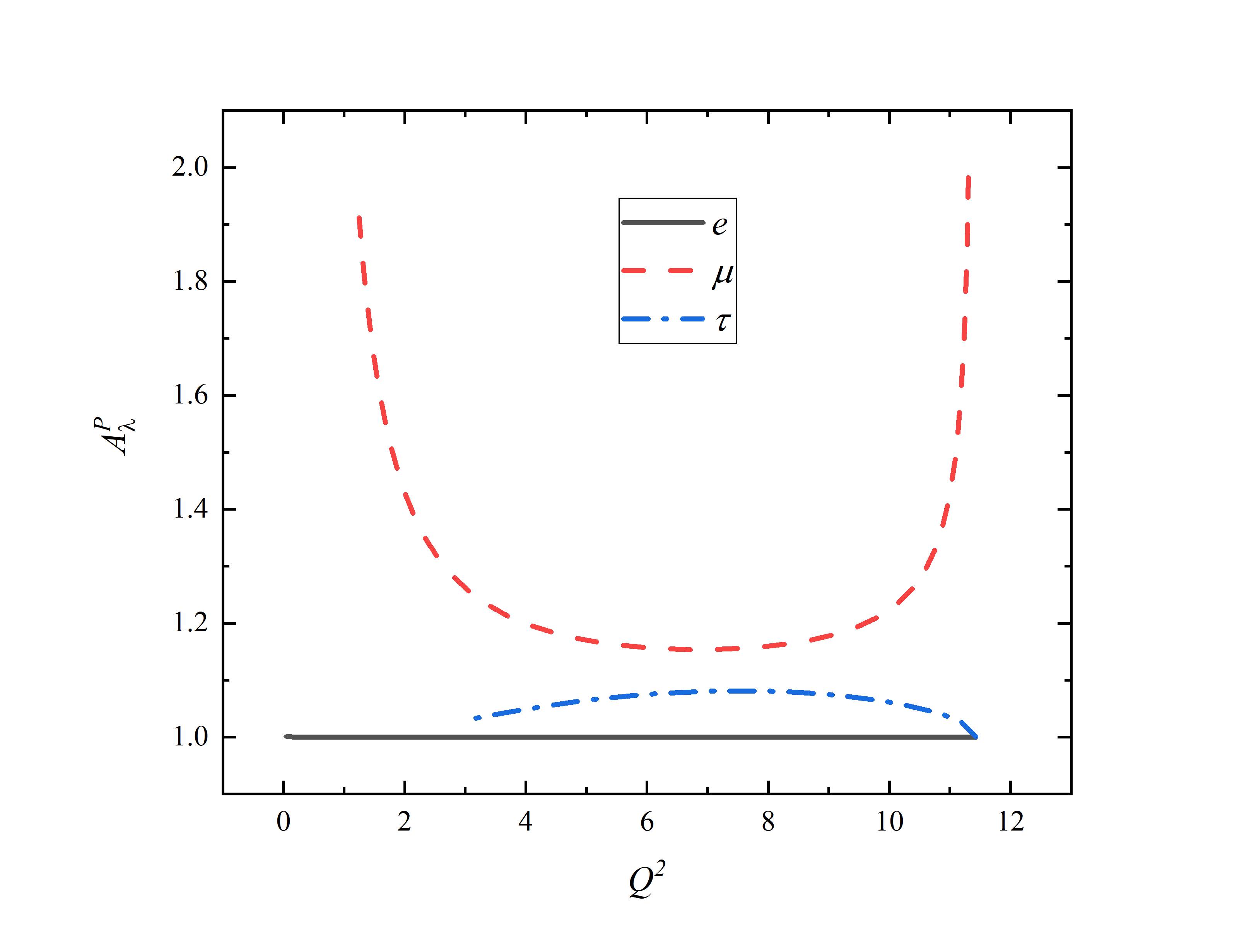}
  \end{minipage}%
  \begin{minipage}[t]{0.5\textwidth}
   \centering
   \includegraphics[width=3.4in]{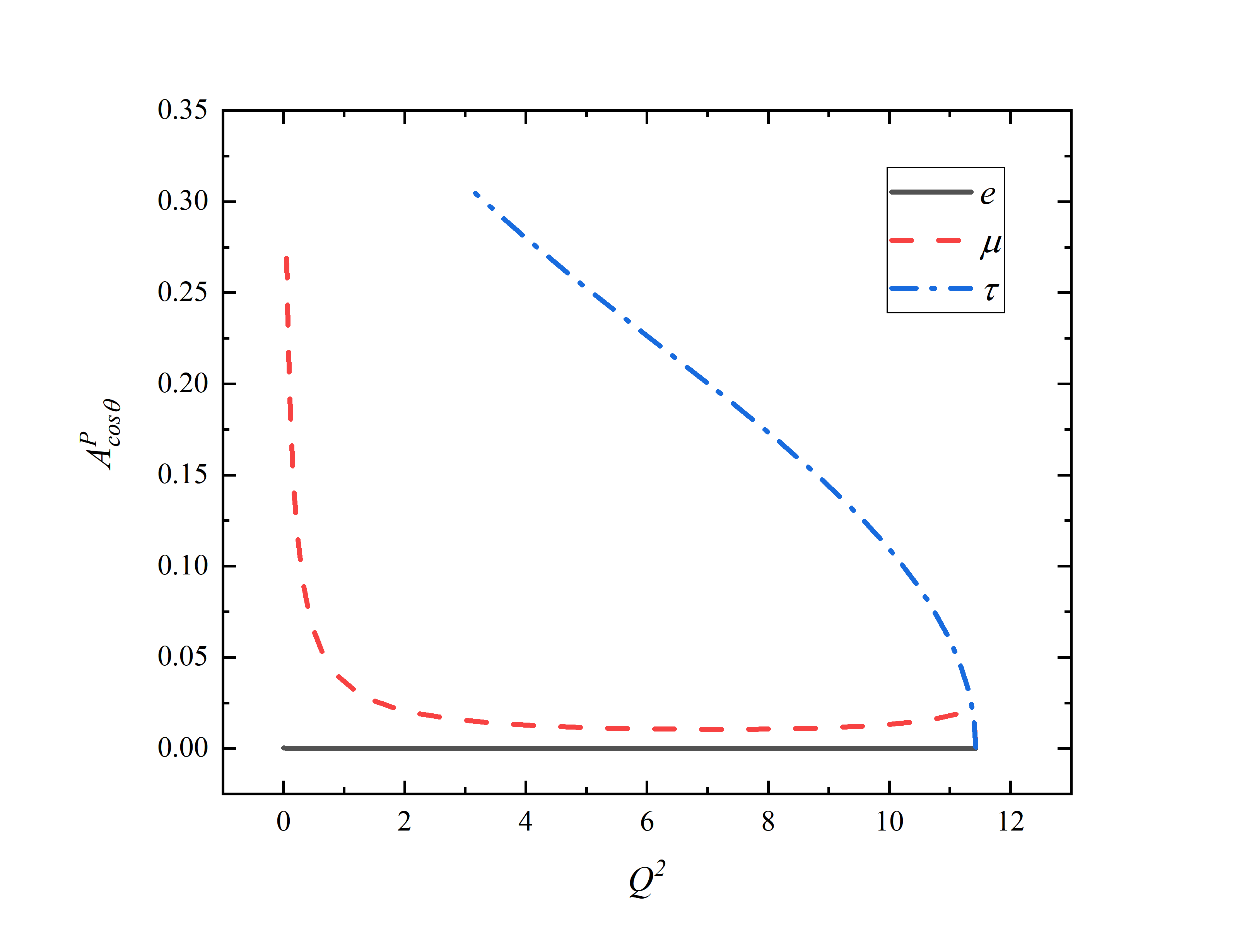}
  \end{minipage}
  \caption{\label{fig:Bsalac} $A_ {\lambda}^{P}$(left) and $A_{cos\theta}^{P}$(right) of the decay $\bar{B}^0_s \rightarrow D_s^{+}(1S)\ell \nu_\ell$.}
\end{figure}

\begin{figure}[h]
  \begin{minipage}[t]{0.5\textwidth}
   \centering
   \includegraphics[width=3.4in]{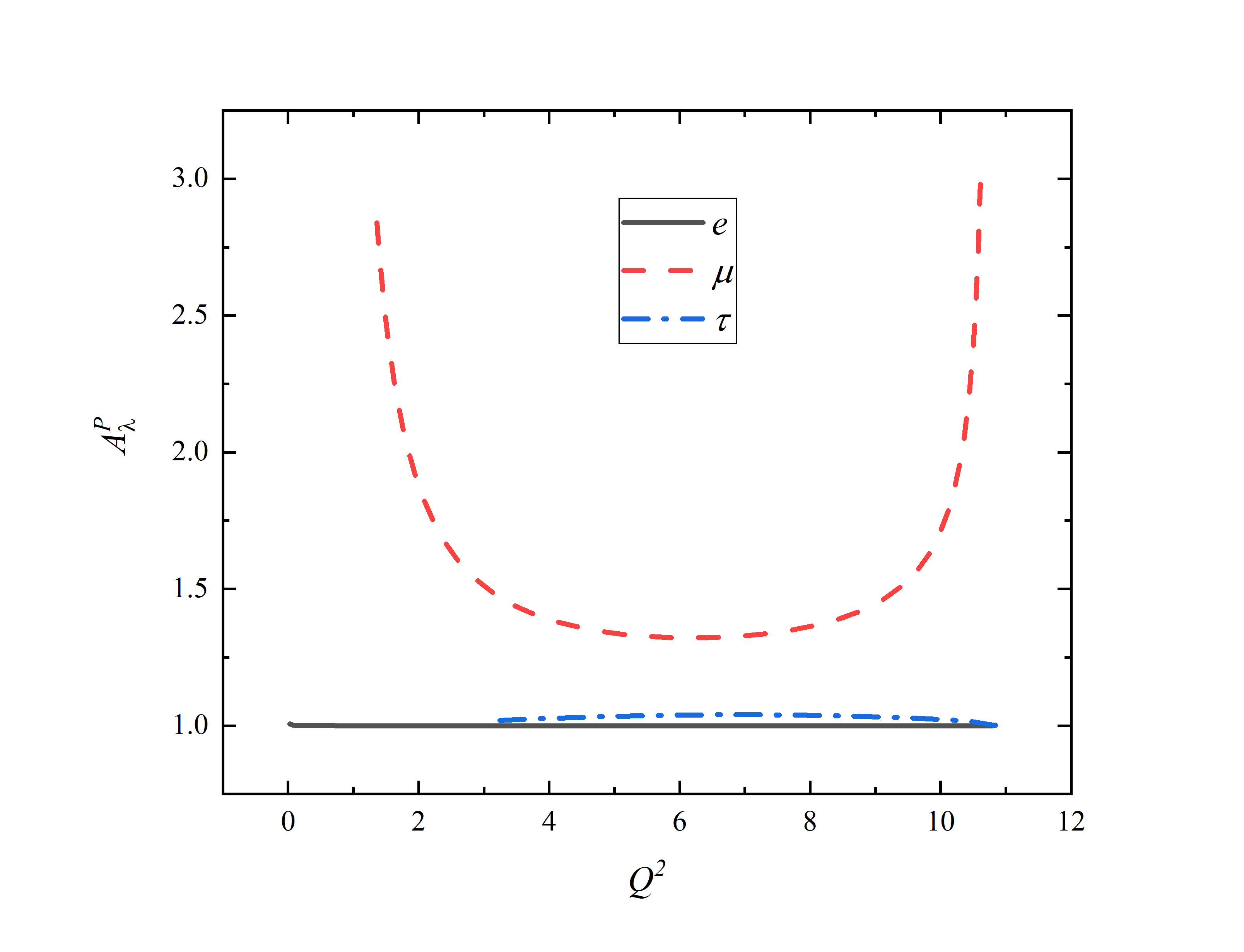}
  \end{minipage}%
  \begin{minipage}[t]{0.5\textwidth}
   \centering
   \includegraphics[width=3.4in]{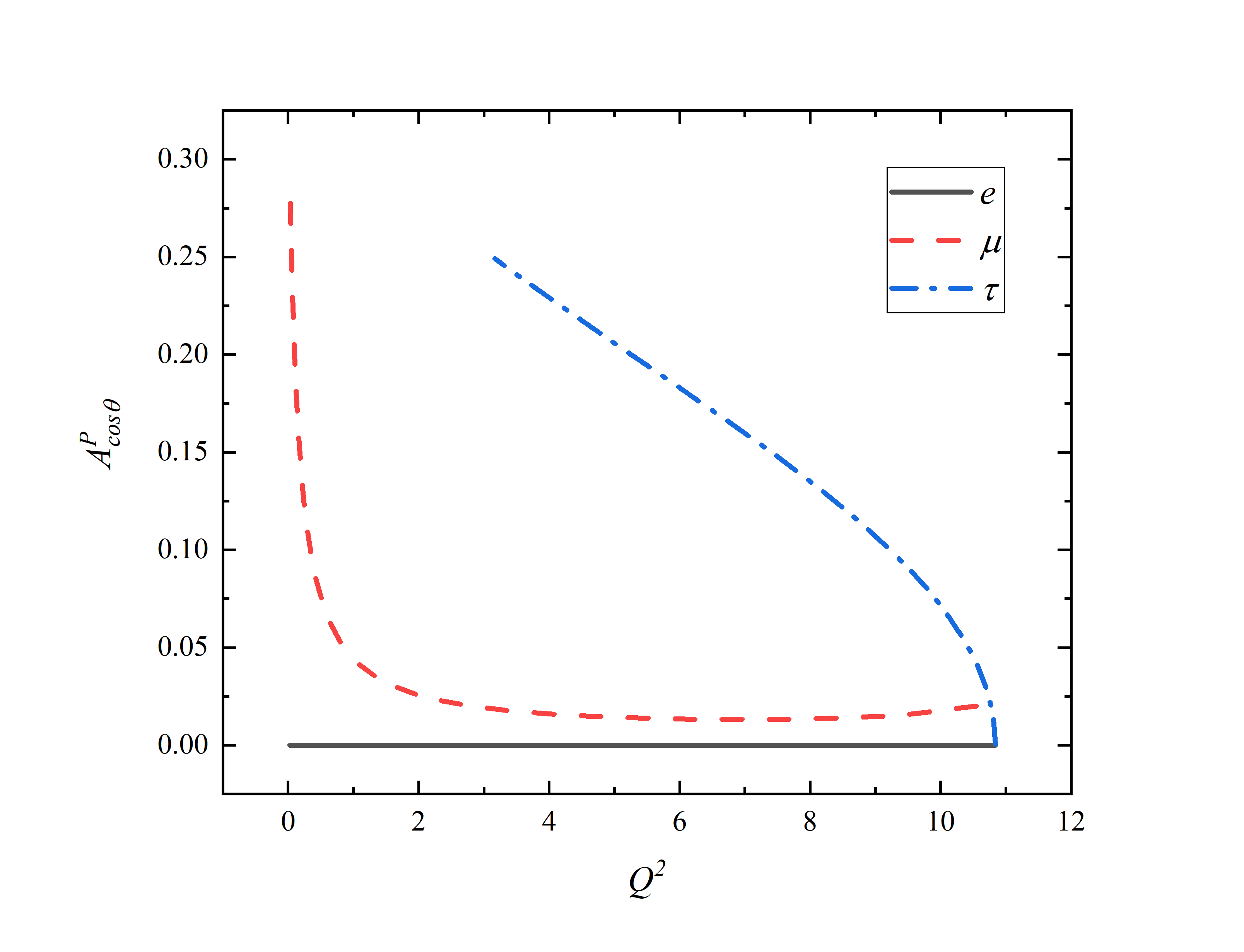}
  \end{minipage}
  \caption{\label{fig:Bcalac} $A_{\lambda}^{P}$(left) and $A_{cos\theta}^{P}$(right) of the decay $B_c^- \rightarrow \eta_c \ell \nu_\ell$.}
\end{figure}

\begin{figure}[h]
  \begin{minipage}[t]{0.5\textwidth}
   \centering
   \includegraphics[width=3.4in]{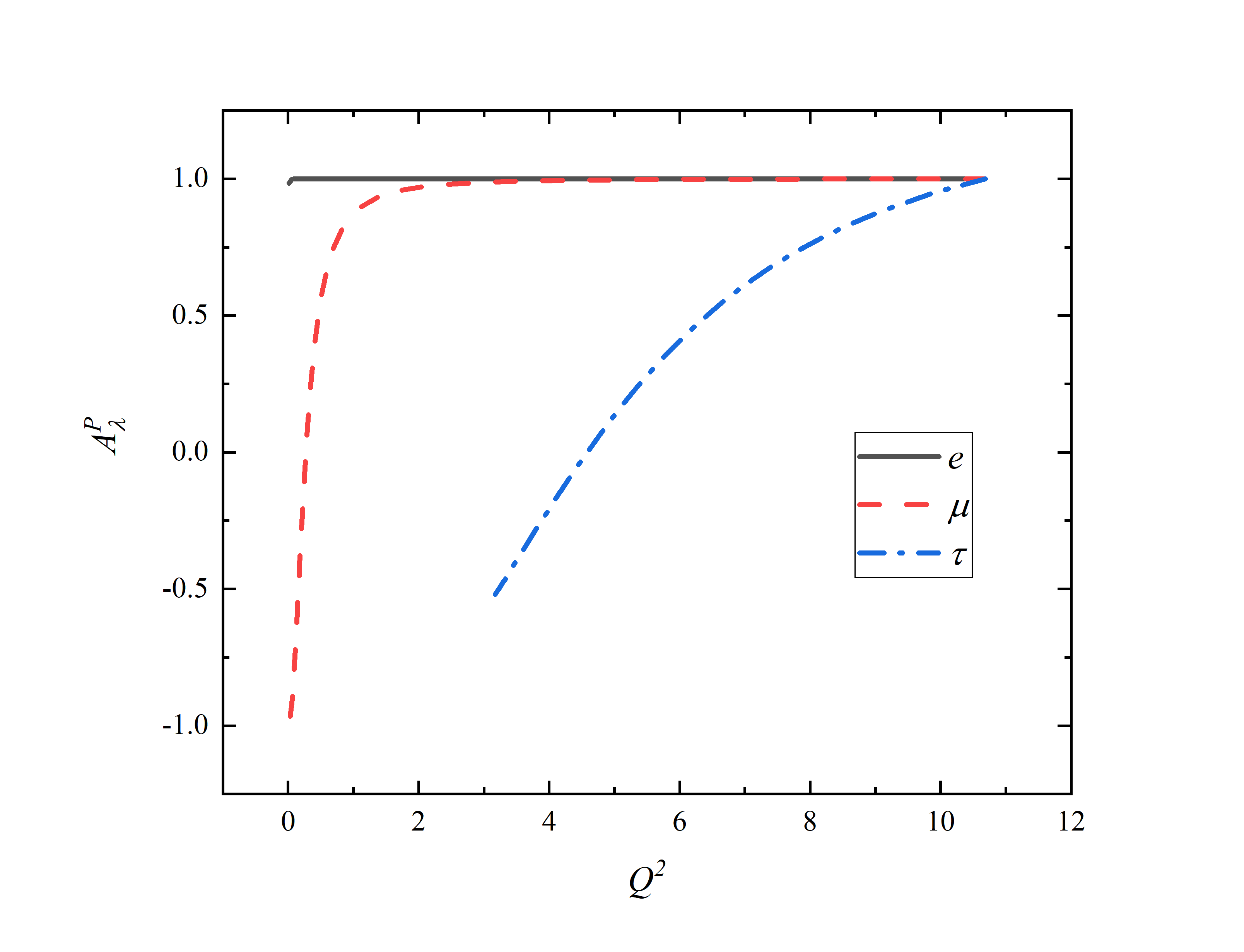}
  \end{minipage}%
  \begin{minipage}[t]{0.5\textwidth}
   \centering
   \includegraphics[width=3.4in]{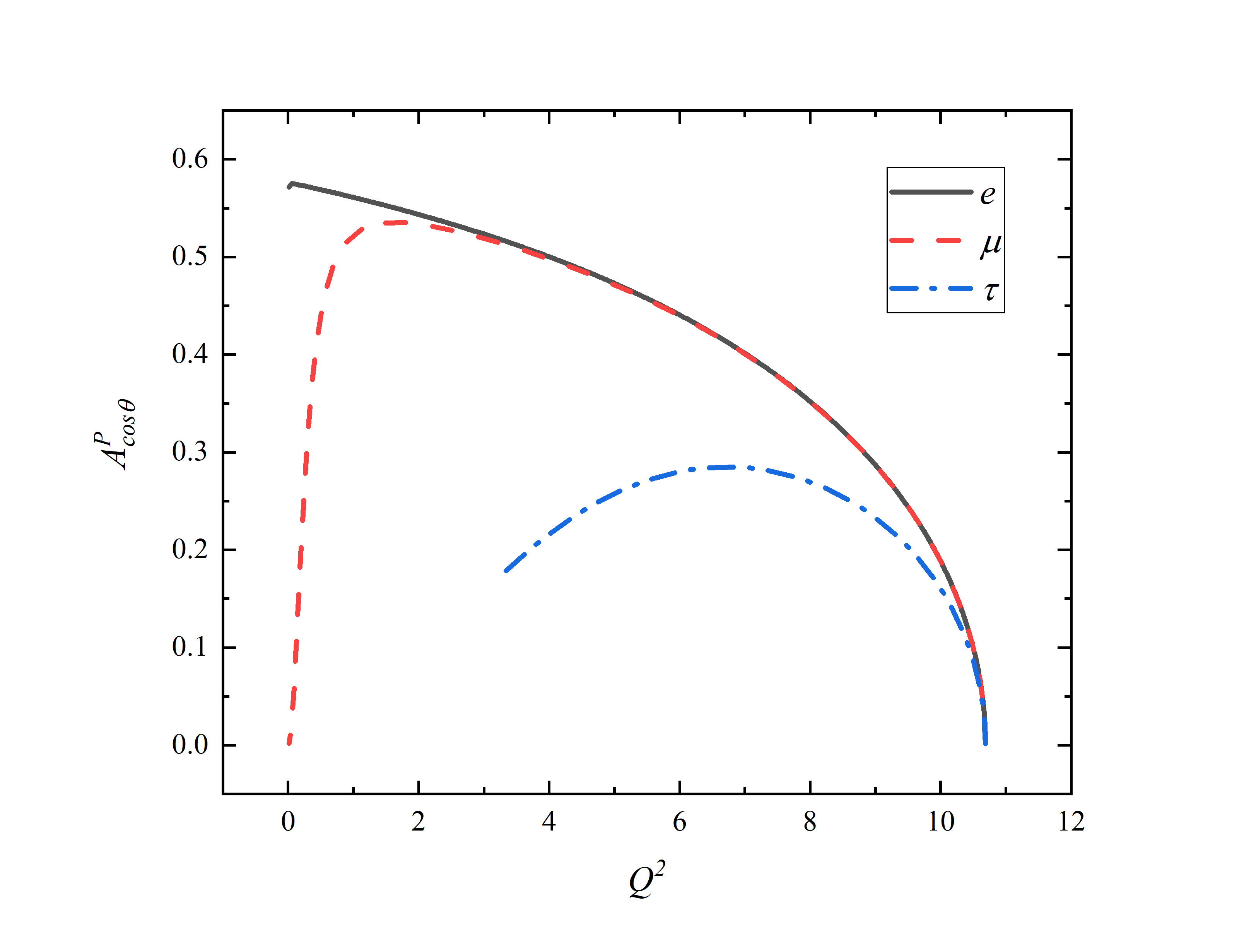}
  \end{minipage}
  \caption{\label{fig:B0salac} $A_{\lambda}^{P}$(left) and $A_{cos\theta}^{P}$(right) of the decay $\bar{B}^0\rightarrow D^{*+}\ell \nu_\ell$}
\end{figure}

\begin{figure}[h]
  \begin{minipage}[t]{0.5\textwidth}
   \centering
   \includegraphics[width=3.4in]{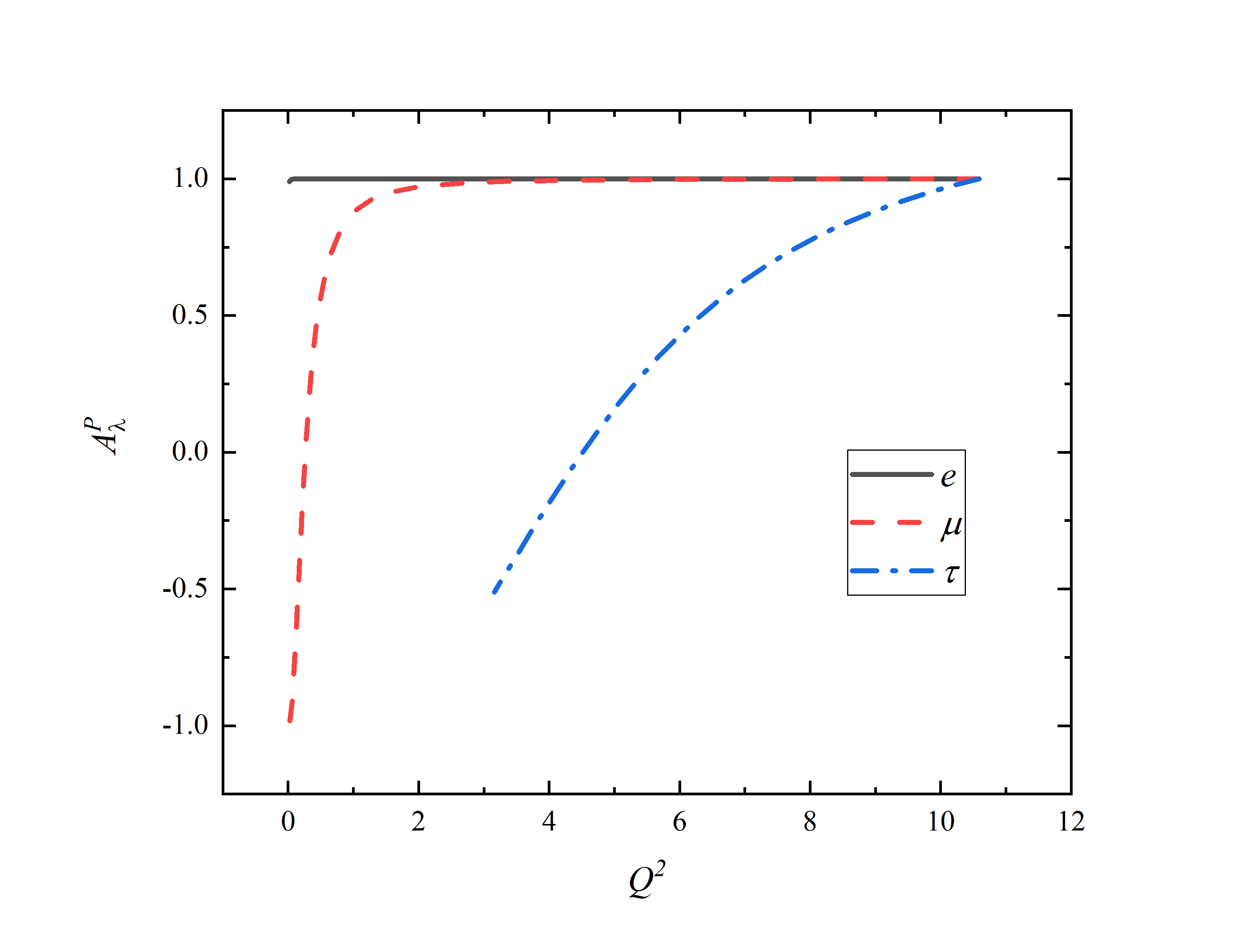}
  \end{minipage}%
  \begin{minipage}[t]{0.5\textwidth}
   \centering
   \includegraphics[width=3.4in]{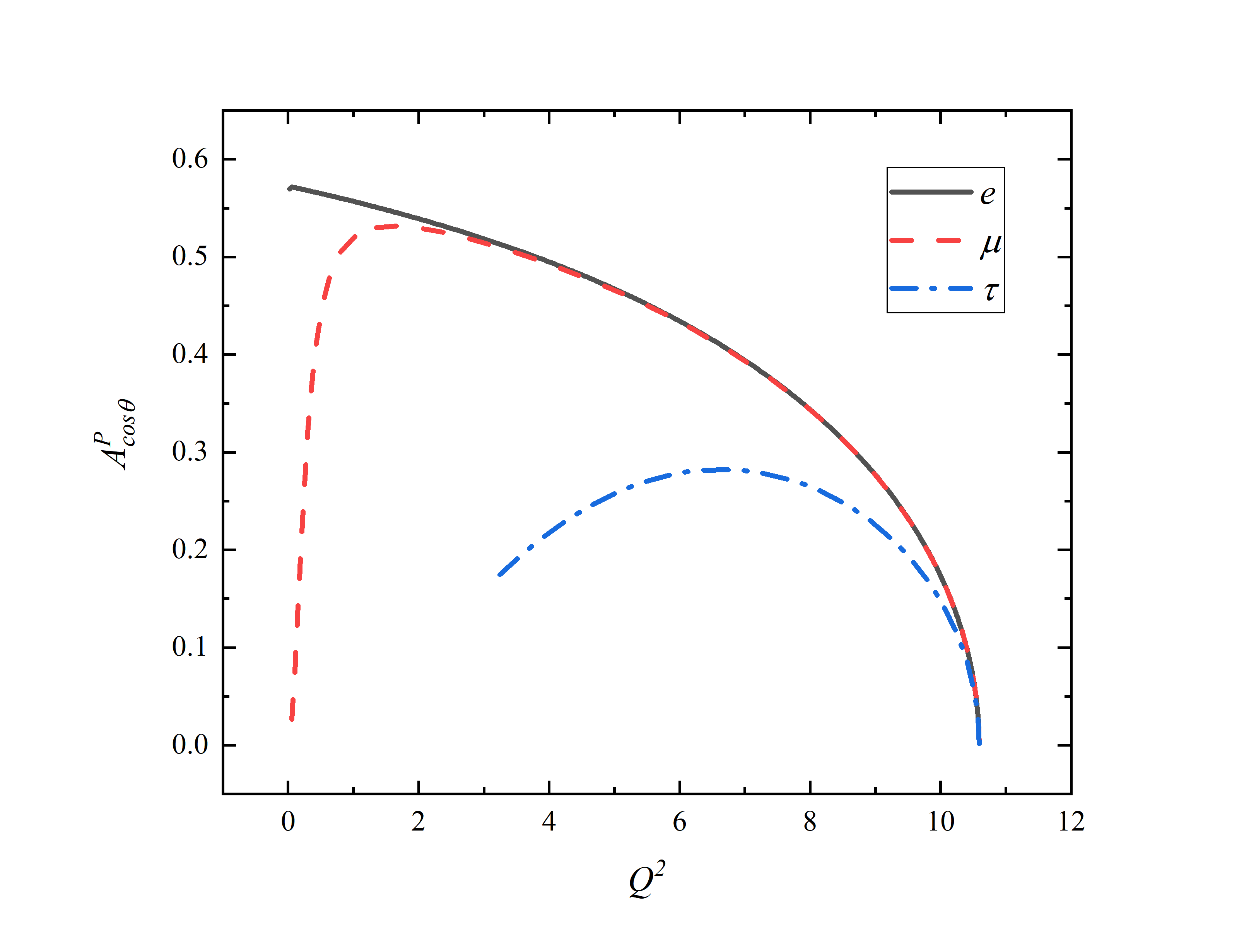}
  \end{minipage}
  \caption{\label{fig:Bssalac} $A_{\lambda}^{P}$(left) and $A_{cos\theta}^{P}$(right) of the decay $\bar{B}^0_s\rightarrow D^{*+}_s(1S)\ell \nu_\ell$.}
\end{figure}

Finally, the decay widths and corresponding branching ratios by the improved BS method are shown in table \ref{tab:1} and table \ref{tab:2}, where  table \ref{tab:1} is the cases of $B_q$ decays to a pseudoscalar and table \ref{tab:2} is for the cases of $B_q$ to a vector. As before, the errors are calculated by varying all the parameters around their center values within $\pm10 \%$.

\begin{table}[htbp]
\centering
 \caption{\label{tab:1} Decay widths ($10^{-15}$ GeV), branching ratios ($\%$) and $R(D)$.}
 \begin{tabular}{ccccccccc}
   \Xhline{1.5pt}
  Channels                          &width      &   Br                  & Br \cite{Polosa:2000ym}  &Br \cite{Azizi:2008tt,Azizi:2008vt} & Br (PDG \cite{Tanabashi:2018oca})               & Ratio ($\tau/\ell$)\\
  \midrule
$B^-\rightarrow D^{0}e\nu_e$      &$7.68^{+2.67}_{-1.84}$      & $1.91^{+0.67}_{-0.45}$  & $2.20-3.00$                  &$1.50-2.40$& $2.20 \pm 0.10$            & $0.312^{+0.006}_{-0.007}$
\\
$B^-\rightarrow D^{0}\mu \nu_\mu$   &$7.65^{+2.65}_{-1.83}$    & $1.91^{+0.66}_{-0.45}$ && &$2.20 \pm 0.10$            & $0.313^{+0.006}_{-0.007}$
\\
$B^-\rightarrow D^{0}\tau \nu_\tau$  &$2.40^{+0.77}_{-0.54}$   & $0.60^{+0.20}_{-0.13}$  &&&$0.77 \pm 0.25$
 \\
\hline
$\bar{B}^0\rightarrow D^{+}e\nu_e$      &$7.65^{+2.67}_{-1.83}$      &  $1.77^{+0.62}_{-0.42}$& $2.20-3.00$                  &$1.30-2.20$           &$2.20 \pm 0.10$            & $0.312^{+0.006}_{-0.006}$
\\
$\bar{B}^0\rightarrow D^{+}\mu \nu_\mu$    &$7.64^{+2.66}_{-1.84}$   & $1.76^{+0.62}_{-0.42}$  &&&$2.20 \pm 0.10$            & $0.313^{+0.006}_{-0.007}$
\\
$\bar{B}^0\rightarrow D^{+}\tau \nu_\tau$  &$2.39^{+0.77}_{-0.54}$   & $0.55^{+0.18}_{-0.12}$  &&&$1.03 \pm 0.22$
\\
\hline
$\bar{B}_s^0\rightarrow D_s^{+}e\nu_e$     &$7.12^{+2.73}_{-1.84}$   & $1.64^{+0.63}_{-0.42}$   & &$2.80-3.80$            &                          & $0.320^{+0.009}_{-0.009}$
\\
$\bar{B}_s^0\rightarrow D_s^{+}\mu \nu_\mu$ &$7.10^{+2.73}_{-1.83}$  & $1.63^{+0.63}_{-0.42}$    & &&                         & $0.321^{+0.009}_{-0.009}$
\\
$\bar{B}_s^0\rightarrow D_s^{+}\tau \nu_\tau$ &$2.28^{+0.80}_{-0.54}$ &  $0.52^{+0.18}_{-0.12}$ &&&
\\
\hline
$B_c^-\rightarrow \eta_ce\nu_e$      &$5.25^{+2.68}_{-1.80}$   & $0.40^{+0.21}_{-0.14}$    & $0.45$ {\cite{Ebert:2003cn}}     &$0.50$ {\cite{Ivanov:2006ni}}& & $0.384^{+0.032}_{-0.042}$
\\
$B_c^-\rightarrow \eta_c\mu \nu_\mu$  &$5.25^{+2.69}_{-1.79}$ & $0.40^{+0.21}_{-0.14}$  && &                         & $0.384^{+0.033}_{-0.041}$
\\
$B_c^-\rightarrow \eta_c\tau \nu_\tau$ &$2.02^{+0.77}_{-0.55}$ & $0.15^{+0.06}_{-0.04}$  &&&
\\
\bottomrule
 \end{tabular}
\end{table}

\begin{table}[htbp]
\centering
 \caption{\label{tab:2} Decay widths ($10^{-14}$ GeV), branching ratios ($\%$) and $R(D^*)$.}
 \begin{tabular}{ccccccccc}
   \Xhline{1.5pt}
  Channels     &width   &  Br     & Br \cite{Polosa:2000ym}  &Br \cite{Azizi:2008tt,Azizi:2008vt}&Br (PDG \cite{Tanabashi:2018oca})              &Ratio($\tau/\ell$) \\
  \midrule
 $B^-\rightarrow D^{*0}e\nu_e$       &$2.62^{+0.75}_{-0.55}$     & $6.54^{+1.88}_{-1.37}$    & $5.90-7.60$ &$4.36-8.94$&$4.88 \pm 0.10$      & $0.249^{+0.001}_{-0.002}$
\\
 $B^-\rightarrow D^{*0}\mu \nu_\mu$    &$2.61^{+0.75}_{-0.55}$   & $6.51^{+1.88}_{-1.36}$    &&&$4.88 \pm 0.10$           & $0.249^{+0.001}_{-0.002}$
\\
 $B^-\rightarrow D^{*0}\tau \nu_\tau$  &$0.65^{+0.18}_{-0.13}$   & $1.63^{+0.48}_{-0.33}$    &&&$1.88 \pm 0.20$
 \\
\hline
 $\bar{B}^0\rightarrow D^{*+}e\nu_e$        &$2.61^{+0.75}_{-0.55}$    & $6.03^{+1.74}_{-1.26}$    & $5.90-7.60$   &$4.57-9.12$ &$4.88 \pm 0.10$           & $0.248^{+0.001}_{-0.002}$
\\
 $\bar{B}^0\rightarrow D^{*+}\mu \nu_\mu$   &$2.60^{+0.75}_{-0.54}$    & $6.01^{+1.74}_{-1.25}$    &&&$4.88 \pm 0.10$           & $0.249^{+0.001}_{-0.002}$
\\
 $\bar{B}^0\rightarrow D^{*+}\tau \nu_\tau$  &$0.65^{+0.18}_{-0.13}$   & $1.50^{+0.42}_{-0.30}$    &&&$1.67 \pm 0.13$
\\
\hline
 $\bar{B}_s^0\rightarrow D_s^{*+}e\nu_e$    &$2.50^{+0.78}_{-0.55}$    & $5.74^{+1.79}_{-1.27}$        &  &  $1.89-6.61$     &$2.80-3.80$      & $0.251^{+0.002}_{-0.003}$
\\
 $\bar{B}_s^0\rightarrow D_s^{*+}\mu \nu_\mu$  &$2.48^{+0.78}_{-0.55}$  & $5.71^{+1.79}_{-1.26}$        &    &&     & $0.252^{+0.002}_{-0.003}$
\\
 $\bar{B}_s^0\rightarrow D_s^{*+}\tau \nu_\tau$ &$0.63^{+0.19}_{-0.13}$ & $1.44^{+0.43}_{-0.31}$   &&   &
\\
\hline
 $B_c^-\rightarrow J/\psi e\nu_e$      &$2.11^{+0.78}_{-0.56}$    & $1.62^{+0.60}_{-0.43}$          &$1.36$ {\cite{Ebert:2003cn}}  &$1.67$ {\cite{Ivanov:2006ni}}  &    & $0.267^{+0.009}_{-0.011}$
\\
 $B_c^-\rightarrow J/\psi\mu \nu_\mu$   &$2.10^{+0.77}_{-0.55}$  & $1.62^{+0.59}_{-0.43}$           &        &&            & $0.268^{+0.009}_{-0.012}$
\\
 $B_c^-\rightarrow J/\psi\tau \nu_\tau$  &$0.56^{+0.18}_{-0.13}$ & $0.43^{+0.14}_{-0.10}$  &&& &
\\
\bottomrule
 \end{tabular}
\end{table}

In table \ref{tab:1} and table \ref{tab:2}, for comparison, we also show other theoretical results as well as the experimental data from PDG. In the last column of table \ref{tab:1}, we show our results of the ratios $R(D)$, $R(D_s)$ and $R(\eta_c)$. And the corresponding vector cases $R(D^*)$, $R(D^*_s)$ and $R(J/\psi)$ are shown in the last column of table \ref{tab:2}. From these two tables, we can see that, our results of branching ratios consist with experimental data within errors . However, for the $B^-\to D^0$ processes, we get a larger value of $R(D)$ than the experimental data. Because the center value of the branching ratio of $B^-\rightarrow D^{0}e\nu_e$ is smaller than that of experimental data, while for the $B^-\rightarrow D^{0}\tau \nu_\tau$ process, the result is opposite.

\begin{table}[htbp]
\centering
 \caption{\label{tab:5} The experiment data and SM prediction of $R(D),R(D^*)$ and $R(J/\psi)$ with the result in this paper.}
 \begin{tabular}{cccc}
   \Xhline{1.5pt}
experiment  &  $R(D)$ & $R(D^*)$ & $R(J/\psi)$ \\
  \midrule
ours & $0.312^{+0.006}_{-0.007}$  & $0.248^{+0.001}_{-0.002}$  & $0.267^{+0.008}_{-0.012}$  \\
Lattice QCD \cite{Harrison:2017fmw}  & $0.300\pm0.008$ & \\
Heavy quark expansion \cite{Fajfer:2012vx} & & $0.252\pm0.003$ & \\
LCSR \cite{Bigi:2017jbd} &  & 0.260(8) & \\
CCQM \cite{Tran:2018kuv} & & & 0.24 \\
BarBar \cite{Amhis:2016xyh} & $0.440\pm0.058\pm0.042$ & $0.332\pm0.024\pm0.018$ \\
Belle (2017) \cite{Amhis:2016xyh} & $0.375\pm0.058\pm0.042$ & $0.293\pm0.038\pm0.015$\\
Belle (2019) \cite{Abdesselam:2019dgh} & $0.307\pm0.037\pm0.016$ &$0.283\pm0.018\pm0.014$\\
LHCb \cite{Amhis:2016xyh,Aaij:2017tyk}&  & $0.285\pm0.019\pm0.029$ & $0.71\pm0.17\pm0.18$ \\
\bottomrule
 \end{tabular}
\end{table}

To compare with each other, we give the table \ref{tab:5}, in which we show the ratios of $R(D)$, $R(D^{(*)})$ and $R(J/\psi)$ by this method, other SM predictions and experimental data. There are many theoretical predictions of the SM available now, while we only present few of them in this paper for comparison. We can see that, though we have relative large theoretical uncertainties in the branching ratios, we get very small uncertainties in the ratios of $R(D)$, $R(D^{(*)})$ and $R(J/\psi)$, as most of the uncertainties are cancelled. This also happens in other theoretical predictions. Our result of $R(D)$ is close to other SM predictions, while larger than  experimental data except the recent Belle data in 2019. For $R(D^{(*)})$, most theoretical results are consistent with each other, but smaller than the experimental data. For $R(J/\psi)$, the theoretical predictions of the SM is much smaller than the experimental data.

In conclusion, we give a relativistic calculation of the ratios $R(D)$, $R(D^*)$ and $R(J/\psi)$ using the instantaneous BS method which has been improved to provide a more covariant formula to calculate the transition matrix element. Within errors, our results of the branching fractions are consistent with the experimental data. However, their ratios $R(D)$, $R(D^*)$ and $R(J/\psi)$, which are consistent with other theoretical predictions, are distinctly different from the experimental data except the recent Belle result \cite{Abdesselam:2019dgh}.

\acknowledgments

This work was supported in part by the National Natural Science
Foundation of China (NSFC) under Grant Nos.~11575048, 11405037, 11505039. We also thank the HEPC Studio at Physics School of Harbin Institute of Technology for access to computing resources through INSPUR-HPC@hepc.hit.edu.cn.

\paragraph{}

\bibliographystyle{JHEP}
\bibliography{bibfile}

\end{document}